\documentstyle[aps,pra,eqsecnum,12pt,preprint,psfig,floats]{revtex}
\tighten
\draft

\begin{document}

\title{Reflection of light from a disordered medium\\ backed by a 
phase-conjugating mirror}
\author{J. C. J. Paasschens,$^{\rm a,b}$
M. J. M. de Jong,$^{\rm a}$
P.  W. Brouwer,$^{\rm b}$ and C. W. J. Beenakker$^{\rm b}$}
\address{$^a$Philips Research Laboratories, 5656 AA Eindhoven, The
      Netherlands\\
$^b$Instituut-Lorentz, Leiden University,
                 P.O. Box 9506, 2300 RA Leiden, The Netherlands}
\date{\today}
\maketitle
\begin{abstract}
This is a theoretical study of the interplay of optical phase-conjugation
and multiple scattering. We calculate the intensity of light reflected by a
phase-conjugating mirror when it is placed behind a disordered medium.
We compare the results of a fully phase-coherent theory with those from
the theory of radiative transfer. Both methods are equivalent if the
dwell time $\tau_{\rm dwell}$ of a photon in the disordered medium is
much larger than the inverse of the  frequency shift $2\Delta\omega$
acquired at the phase-conjugating mirror. When $\tau_{\rm dwell}\Delta\omega
\lesssim 1$, in contrast, phase coherence drastically affects the
reflected intensity. In particular, a minimum in the dependence of the 
reflectance on
the disorder strength disappears when $\Delta\omega$ is reduced below
$1/\tau_{\rm dwell}$. The analogies and differences with Andreev
reflection of electrons at the interface between a normal metal and
a superconductor are discussed.
\end{abstract}
\pacs{PACS numbers: 42.65.Hw, 42.68.Ay, 42.25.Bs, 78.20.Ci\\
      \texttt{physics/9708010}}

\section{Introduction}
\label{SecIntro}
Phase conjugation is the reversal of the sign of the phase of a wave
function. A phase-conjugated wave retraces the path of the original
wave, thereby canceling all accumulated phase shifts. Phase conjugation
was first discovered for electronic waves\cite{Andreev64}, and later for
optical waves\cite{Woerdman70,Stepanov71}. For electrons, phase
conjugation takes place at the interface between a normal metal and a
superconductor. An electron at energy $E$ above the Fermi energy $E_F$
is reflected at the angle of incidence (retro-reflected) as a hole at
energy $E$ below $E_F$, a process known as Andreev
reflection\cite{Abrikosov}.
A phase-conjugating mirror for light consists of a cell containing a
liquid or crystal with a large nonlinear susceptibility, pumped by two
counter-propagating beams at frequency $\omega_0$. A wave incident at
frequency $\omega_0+\Delta\omega$ is then retro-reflected at frequency
$\omega_0-\Delta\omega$, a process known as four-wave
mixing\cite{Fisher,Zeldovich,Pep86}.

The interplay of multiple scattering by disorder and phase conjugation
has been studied extensively in the electronic case, both experimentally
and theoretically. (See Ref.~\cite{LesHouches} for a review.) In the
optical case the emphasis has been on weakly disordered media, which do
not strongly scatter the waves\cite{Wolf}. 
%(An exception is formed by Refs.~\cite{Mittra84,Krav90}.)
Complete wave-front reconstruction is possible
only if the distorted wave front remains approximately planar,
since perfect time reversal upon reflection 
holds only in a narrow range of angles of incidence
  for realistic systems. (For the hypothetical case of perfect
  time-reversal at all angles, see Ref.~\cite{Mittra84}.)
McMichael, Ewbank, and Vachss \cite{McMichael95}
  measured the intensity of the reconstructed wave front for a
  strongly inhomogeneous medium (small transmission probability
  $T_{0}$), and found that it was proportional to $T_{0}^{2}$ --- in
  agreement with the theoretical prediction of Gu and Yeh \cite{Gu94}.
If $T_{0}\ll 1$, the intensity  of the reconstructed wave is much
  smaller than the total reflected intensity. The total reflected
  intensity was not studied previously, perhaps because it was believed
  that the diffusive illumination resulting from a strongly
  inhomogeneous medium would render the effect of phase conjugation
  insignificant.
In this paper we show that a strongly
disordered medium backed by a phase-conjugating mirror has unusual
optical properties, different both from the weakly disordered case and
{}from the electronic analogue.

We distinguish two regimes, depending on the relative magnitude of the
frequency shift $2\Delta\omega$ acquired at the phase-conjugating mirror
and the inverse of the dwell time $\tau_{\rm dwell}$ of a photon in the
disordered medium. (For a medium of length $L$ and mean free path $l$,
with light velocity $c$, one has $\tau_{\rm dwell}\simeq L^2/cl$.) In
the {\it coherent regime}, $\Delta\omega\ll 1/\tau_{\rm dwell}$, phase
conjugation leads to a constructive interference of multiply scattered
light in the disordered medium. In the {\it incoherent regime},
$\Delta\omega\gg1/\tau_{\rm dwell}$, interference effects are
insignificant. In both regimes we compute the reflectances $R_+$ and
$R_-$, defined as the reflected power at frequency
$\omega_0\pm\Delta\omega$ divided by the incident power at frequency
$\omega_0+\Delta\omega$. A distinguishing feature of the two regimes is
that (in a certain parameter range) the reflectance $R_-$ decreases
monotonically as a function of $L/l$ in the coherent regime, while in
the incoherent regime it first decreases and then increases.

The outline of this paper is as follows. After having formulated the
problem in Sec.~\ref{SecFormulation}, we discuss in
Sec.~\ref{SecIncoherent} its solution using the Boltzmann equation,
ignoring phase coherence. This is the theory of radiative 
transfer\cite{Chan,Ish77}. A simple result is obtained 
if we neglect angular correlations between the
scattering in the disordered medium and at the phase-conjugating mirror. 
We compare this approximation with an exact solution of the
Boltzmann equation. In Sec.~\ref{SecPhase} the phase-coherent
problem is addressed, analytically using random-matrix theory, and
numerically using the method of recursive Green functions. Results of
this section were briefly presented in Ref.~\cite{Paa96}.
We conclude in Sec.~\ref{SecConclusions} with a comparison with the
electronic analogue of this problem.

\section{Formulation of the problem}
\label{SecFormulation}
\begin{figure}
\centerline{\psfig{file=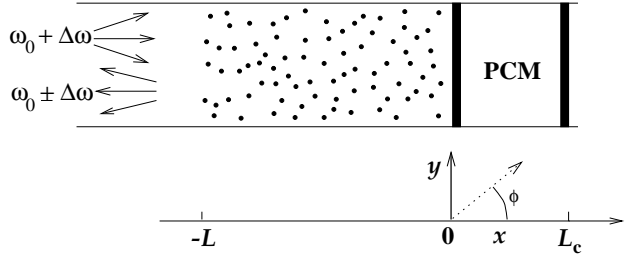,angle=270,width=0.5\hsize}}
\medskip
\caption{Schematic drawing of the disordered medium backed by a
phase-conjugating mirror (PCM). Light incident at frequency
$\omega_0+\Delta\omega$ is reflected at the two frequencies
$\omega_0\pm\Delta\omega$.}
\label{system}
\end{figure}

We study the system shown in Fig.~\ref{system}. It consists
of a disordered medium (length $L$, mean free path $l$), backed at one
end by a phase-conjugating mirror.
The other end is illuminated diffusively
at frequency $\omega_+ = \omega_0+\Delta\omega$, where $\omega_0$ is the
pump frequency of the mirror.
We are interested in the amount of light reflected at frequency
$\omega_+$ and $\omega_-=\omega_0-\Delta\omega$.

To reduce the problem to the scattering of a {\em scalar} wave, we
choose a two-dimensional geometry. The scatterers consist of dielectric
rods in the $z$-direction, randomly placed in the $x$-$y$-plane. The
electric field points in the $z$-direction and varies in the
$x$-$y$-plane only. Two-dimensional scatterers are somewhat artificial,
but can be realized experimentally\cite{Freund88}. We believe that our
results apply qualitatively to a three-dimensional geometry as well,
because the randomization of the polarization by the disorder renders
the vector character of the light insignificant.

The $z$-component of the electric field at the frequencies $\omega_+$
and $\omega_-$ is given by
\begin{equation}
  E_\pm(x,y,t) = {\rm Re}\; {\cal E}_\pm(x,y) \exp(-i\omega_\pm t).
\end{equation}
The phase-conjugating mirror (at $x=L$) couples the two frequencies via
the wave equation\cite{Fisher,Lenstra,Hout91}
\begin{equation}
  \pmatrix{{\cal H}_0&\gamma^*\cr-\gamma&-{\cal H}_0}\pmatrix{{\cal E}_+\cr
   {\cal E}_-^*} =
  \frac{2\varepsilon\Delta\omega}{\omega_0}\,\pmatrix{{\cal E}_+\cr {\cal E}_-^*}.
   \label{BdGlike}
\end{equation}
The complex dimensionless coupling constant $\gamma$ is zero for $x<0$
and for $x>L_c$, with $L_c$ the length of the nonlinear medium
forming the phase-conjugating mirror. For $0<x<L_c$ it is 
proportional to the electric
fields ${\cal E}_1$, ${\cal E}_2$ of the two pump beams and to the third-order nonlinear
susceptibility $\chi_3$: 
\begin{equation}
  \gamma = - \frac3{2\varepsilon_0} \chi_3{\cal E}_1^*{\cal E}_2^*
  \equiv \gamma_0 e^{i\psi},\qquad 0<x<L_c.
  \label{gammadef}
\end{equation}
The Helmholtz operator ${\cal H}_0$ at frequency $\omega_0$ is given by
\begin{equation}
   {\cal H}_0 =
   -k_0^{-2} \nabla^2 -\varepsilon,
  \label{Hdef}
\end{equation}
where $\varepsilon(x,y)$ is the relative dielectric constant of the
medium. We take $\varepsilon=1$ except in the disordered region $-L<x<0$,
where $\varepsilon = 1 + \delta\varepsilon(x,y)$. The fluctuations
$\delta\varepsilon$ lead to scattering with mean free path $l$. We
assume $k_0l\gg 1$, where $k_0=\omega_0/c$ is the wave number of the
light (velocity $c$).
The validity of Eq.~(\ref{BdGlike}) requires $\Delta\omega/\omega_0\ll1$
and $|\gamma|\equiv \gamma_0 \ll 1$. The ratio of these two small parameters 
\begin{equation}
  \delta= \frac{2\Delta\omega}{\gamma_0\omega_0}
  \label{deltadef}
\end{equation}
is a measure of the degeneracy of the incident and the reflected
wave, and can be chosen freely.

\begin{figure}
\centerline{\psfig{file=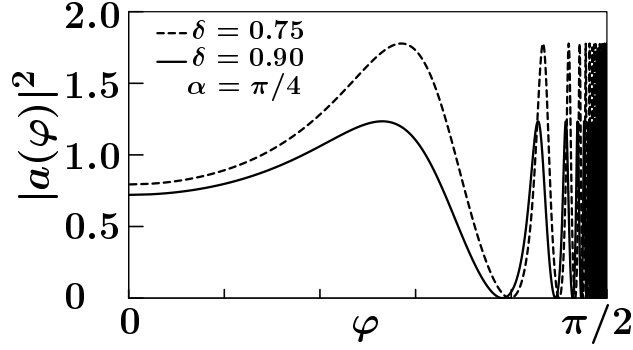,width=0.5\hsize}}
\medskip
\caption{Reflectance of the phase-conjugating mirror as
function of the angle of incidence, computed from
Eq.~(\protect\ref{adeffull}) for  two choices of parameters.}
\label{aphi}
\end{figure}

In the absence of disorder, an incoming plane wave in the direction
$(\cos\phi,\sin\phi)$ is retro-reflected in the direction
$(-\cos\phi,-\sin\phi)$, with a different frequency and amplitude. The
scattering matrix for retro-reflection is given by
\onlinecite{Fisher,Lenstra,Hout91,Yariv77,Arnold89}
\begin{mathletters}
\label{adeffull}
\begin{eqnarray}
  \pmatrix{{\cal E}_+\cr {\cal E}_-^*}^{\rm out} &=&
  \pmatrix{0  & -i a(\phi) e^{-i\psi} \cr i a(\phi) e^{i\psi} & 0}
  \pmatrix{{\cal E}_+\cr {\cal E}_-^*}^{\rm in} ,
  \label{PCMscatmat}\\
  a(\phi) &=& [
     \sqrt{1+\delta^2}{\rm cotan}\ (\alpha\sqrt{1+\delta^2}/\cos\phi)
      +i\delta]^{-1},
  \label{adef}\\
  \alpha &=& \case12 \gamma_0 k_0 L_c .
  \label{alpha}
\end{eqnarray}
\end{mathletters}%
The crucial difference with Ref.~\cite{Mittra84} is that the
reflectance is angle dependent and that the reflection matrix is
non-Hermitian. This implies that not all phases will
be canceled in the conjugation process.
In Fig.~\ref{aphi} we have plotted the reflectance $|a|^2$ as a function of 
the angle of
incidence $\phi$ for $\alpha =\pi/4$ and two values of~$\delta=0.75$ and
$0.9$.
The value $\alpha=\pi/4$ is chosen such that $a=1$ for normal incidence
at frequency $\omega_0$ (i.e.\ for $\phi=0$ , $\delta=0$). The two
values of $\delta$ have been chosen such that the angular average of
the reflectance,
\begin{equation}
  A =\int_0^{\pi/2} d\phi\,\cos\phi\;|a(\phi)|^2,
  \label{Adef}
\end{equation}
is $>1$ for $\delta=0.75$ and $<1$ for $\delta=0.9$. (The $\cos\phi$ weight
factor in Eq.~(\ref{Adef})
corresponds to diffusive illumination.) In most of the numerical
examples throughout this paper we will use these values of $\alpha$ and
$\delta$.

\section{Phase-incoherent solution}
\label{SecIncoherent}
\subsection{Radiative transfer theory}
\label{SSecTransfer}
Within the framework of radiative transfer theory\cite{Chan,Ish77},
the stationary distribution $I(x,y,\phi)\propto|{\cal E}|^2$ of the light 
intensity, at
frequency $\omega$ and wavevector $(k\cos\phi,k\sin\phi)$,
is governed by the Boltzmann equation
\begin{equation}
   \left(l\cos\phi\,\frac{\partial}{\partial x} +
   l\sin\phi\,\frac{\partial}{\partial y} \right)I(x,y,\phi)=
  -I(x,y,\phi) +
   \frac1{2\pi} \int_0^{2\pi} d\phi'\, I(x,y,\phi') .
  \label{boltz}
\end{equation}
We neglect absorption and assume isotropic scattering 
in the $x$-$y$ plane, with
mean free path~$l$. The phase-conjugating mirror couples the intensities
$I_\pm$ of light at the two frequencies $\omega_\pm = \omega_0\pm\Delta\omega$.
We assume that $l$ is independent of frequency.
The symmetry of the system implies that $I(x,y,\phi)=I(x,|\phi|)$.
In this section we take $\phi\in[0,\pi]$. 
For each frequency the Boltzmann equation takes the form
\begin{mathletters}
  \label{boltz2}
\begin{eqnarray}
  l\cos\phi\frac{\partial I_\pm(x,\phi)}{\partial x} &=& \bar I_\pm(x) -
     I_\pm(x,\phi),\\
  \bar I_\pm(x) &=& \frac1\pi \int_0^\pi d\phi\, I_\pm(x,\phi).
\end{eqnarray}
\end{mathletters}%

Eq.~(\ref{boltz2}) has to be supplemented by boundary conditions at
the two ends $x=-L$ and $x=0$ of the disordered medium. 
We consider a situation that 
the system is illuminated at $x=-L$ with diffusive light at 
frequency~$\omega_+$, hence 
\begin{mathletters}
  \label{BC1}
\begin{eqnarray}
  &I_+(-L,\phi) = I_0,\qquad &\mbox{for $\cos\phi>0$,}\\
  &I_-(-L,\phi) = 0,  \;\qquad &\mbox{for $\cos\phi>0$.}
\end{eqnarray}
\end{mathletters}%
At $x = 0$ the light is reflected by the  phase-conjugating mirror. The
intensity is multiplied by
\begin{equation}
  |a(\phi)|^2 = 
       \frac{\sin^2(\alpha\sqrt{1+\delta^2}/\cos\phi)}
       {\delta^2 + \cos^2(\alpha\sqrt{1+\delta^2}/\cos\phi)},
  \label{gdef}
\end{equation}
according to Eq.~(\ref{adeffull}). The reflection is accompanied by a
change in frequency $\omega_\pm\to\omega_\mp$,
so that the boundary condition is
\begin{equation}
  I_\pm(0,\phi) = |a(\phi)|^2\, I_\mp(0,\pi-\phi),
  \qquad\mbox{for $\cos\phi<0$.}
  \label{BC2}
\end{equation}

The flux $j_\pm$ associated with the intensity $I_\pm$ is defined by
\begin{equation}
  j_\pm = \int_0^\pi d\phi\,\cos\phi\,I_\pm(x,\phi),
  \label{jdef}
\end{equation}
and is independent of $x$ [$\partial j_\pm/\partial x=0$ according to
Eq.~(\ref{boltz2})]. The reflectance $R_-$ is defined as the
ratio of the outgoing flux at frequency $\omega_-$ and
the incoming flux at frequency~$\omega_+$,
\begin{equation}
  R_- = -j_-/I_0.
\label{Rmindef}
\end{equation}
The total outgoing flux is $(R_-+R_+)I_0$, where 
\begin{equation}
  R_+=1-j_+/I_0 
\label{Rplusdef}
\end{equation}
is the ratio of the outgoing flux and the incoming flux at the same
frequency $\omega_+$.

\subsection{Neglect of angular correlations}
\label{SSecNeglect}
A simple analytical treatment is possible if the angular correlations
between multiple reflections by the disorder and the
phase-conjugating mirror are neglected. Here we present this simplified
treatment, and in the next subsection we compare with an exact numerical
solution of the Boltzmann equation.

We first consider the disordered region by itself. The plane-wave
transmission probability $|t(\phi)|^2$ is the ratio of transmitted to
incident flux when the incident light is 
a plane wave in the direction 
$(\cos\phi, \sin\phi)$. The transmission probability $T$ for diffusive
illumination is then given by
\begin{equation}
  T= \int_0^{\pi/2}d\phi\,\cos\phi\,|t(\phi)|^2,
  \label{Tdef}
\end{equation}
such that $T$ is the fraction of the flux incident from a diffusive
source which is transmitted through the disordered region. This
probability has been calculated in Ref.~\cite{Jong94} from the
Boltzmann equation~(\ref{boltz2}). The result is
\begin{equation}
  T = \left(1+2\eta L/\pi l\right)^{-1},
  \label{Tinterpol}
\end{equation}
where $\eta$ is a numerical coefficient which depends 
weakly on $L/l$.
In the ballistic limit ($L/l\to0$) $\eta$ has the value $\pi^2/8$ and
in the diffusive limit ($L/l\to\infty$) $\eta$~equals~1. In this
subsection (but not in the next) we
take $\eta=1$ for all $L/l$ for simplicity. 

We use Eq.~(\ref{Tinterpol}) to obtain
the reflectance $R_\pm$ for the case that the disordered
medium is backed by a
phase-conjugating mirror with reflectance
\begin{equation}
  A = \int_0^{\pi/2} d\phi\;\frac{\cos\phi\,\sin^2(\alpha\sqrt{1+\delta^2}/\cos\phi)}
                                 {\delta^2+\cos^2(\alpha\sqrt{1+\delta^2}/\cos\phi)}.
\end{equation}
Since $T$ and $A$ are angular averages, we are
neglecting angular correlations.
The light that comes out at frequency $\omega_-$ has been reflected
an odd number of times at the mirror. The light that has been reflected
once has traversed the medium twice, which leads to a contribution
$T^2A$ to $R_-$. Light that has been reflected three times by the mirror
contributes $T^2A^3(1-T)^2$, since it has been reflected two times by the
medium (each time with probability $1-T$). Summing all contributions,
one finds
\begin{mathletters}
  \label{ballminplus}
\begin{equation}
  R_- = T^2A + T^2A^3(1-T)^2 + T^2A^5(1-T)^4 + \cdots =
        \frac{T^2A}{1-(1-T)^2A^2}.
  \label{ballmin}
\end{equation}
Light that comes out at frequency $\omega_+$ has been
reflected an even number of times at the mirror. Zero reflections by the
mirror contributes $1-T$ to $R_+$, two reflections contributes $T^2A^2(1-T)$,
and four reflections $T^2A^4(1-T)^3$. Summing the series, one finds
\begin{equation}
  R_+ = 1-T + \frac{T^2(1-T)A^2}{1-(1-T)^2A^2} .
  \label{ballplus}
\end{equation}
\end{mathletters}%

The geometric series leading to Eq.~(\ref{ballminplus})
diverges if $(1-T)A\ge1$. This indicates that there is
only a stationary solution to the Boltzmann equation if both the gain at
the mirror and the scattering in the medium are sufficiently weak. If
$A$ is increased at fixed $\alpha=\pi/4$ by reducing $\delta$, the
reflectances $R_\pm$ diverge when $\delta=\delta_c$. 
(This divergence is preempted by depletion of the pump beams in
the phase-conjugating mirror, which we do not describe.)
In the approximation of this subsection, $\delta_c$ is determined 
by $(1-T)A=1$, or $L/l = \case12\pi (A-1)^{-1}$.
In the ballistic limit, $T=1$ and $A<\infty$ for any $\delta>0$.
In the diffusive limit, $T=0$ and $A=1$ for $\delta=0.78$. Hence,
$\delta_c$ increases from 0 to $0.78$ as $L/l$ increases from 0 to $\infty$.

\subsection{Exact solution of the Boltzmann equation}
\label{SSecExact}

The Boltzmann equation~(\ref{boltz2}) can be solved exactly, without
neglect of angular correlations, by adapting the method of
Ref.\cite{Jong94} to an angle-dependent boundary condition.
We first rewrite Eq.~(\ref{boltz2}) as 
\begin{equation}
  \frac{\partial}{\partial x} e^{x/l\cos\phi} I_\pm(x,\phi) =
    \frac1{l\cos\phi} e^{x/l\cos\phi}\bar I_\pm(x),
\end{equation}
and then integrate once over $x$, using the boundary
conditions~(\ref{BC1}) and~(\ref{BC2}). The result is
\begin{mathletters}
\label{Formalsol1}
\begin{eqnarray}
  \label{iplusa}
  I_+(x,\phi) &=
     \displaystyle\hphantom{-}
     \int_{-L}^x \frac{dx'}{l\cos\phi}\, e^{-(x-x')/l\cos\phi}\,
     \bar I_+(x') + I_0e^{-(L+x)/l\cos\phi}, 
\egroup$\hfill$\bgroup
  \;\;&\mbox{for $\cos\phi>0$,}\\
  \label{iplusb}
  I_+(x,\phi) &=\displaystyle
     -\int_x^0 \frac{dx'}{l\cos\phi}\, e^{-(x-x')/l\cos\phi}\,
     \bar I_+(x') 
\egroup$\hfill$\bgroup
  &\nonumber\\
  &\qquad\qquad\qquad\displaystyle
      +e^{-x/l\cos\phi}|a(\phi)|^2\, I_-(0,\pi-\phi), 
\egroup$\hfill$\bgroup
  &\mbox{for $\cos\phi<0$},\\
  \label{imina}
  I_-(x,\phi) &= \displaystyle\hphantom{-}
     \int_{-L}^x \frac{dx'}{l\cos\phi}\, e^{-(x-x')/l\cos\phi}\, \bar I_-(x'), 
\egroup$\hfill$\bgroup
     &\mbox{for $\cos\phi>0$},\\
  \label{iminb}
  I_-(x,\phi) &= \displaystyle
     -\int_x^0 \frac{dx'}{l\cos\phi}\, e^{-(x-x')/l\cos\phi}\,
     \bar I_-(x') 
\egroup$\hfill$\bgroup
  &\nonumber\\
  &\qquad\qquad\qquad\displaystyle
     +e^{-x/l\cos\phi}|a(\phi)|^2\, I_+(0,\pi-\phi), 
\egroup$\hfill$\bgroup
  &\mbox{for $\cos\phi<0$}.
\end{eqnarray}
\end{mathletters}%
Substitution of Eqs.~(\ref{imina}) and~(\ref{iplusa}) into,
respectively, Eqs.~(\ref{iplusb}) and~(\ref{iminb}) yields
\begin{mathletters}
\label{Formalsol2}
\begin{eqnarray}
  I_+(x,\phi) &=&\displaystyle
     \int_x^0 \frac{dx'}{l|\cos\phi|}\, e^{-(x'-x)/l|\cos\phi|}\,
     \bar I_+(x') + e^{x/l|\cos\phi|}|a(\phi)|^2\,
     \nonumber\\
  &&\qquad\displaystyle
     \times\int_{-L}^0\frac{dx'}{l|\cos\phi|}\, e^{x'/l|\cos\phi|} \bar
           I_-(x') , \qquad\qquad\qquad\qquad\qquad\mbox{for $\cos\phi<0$},
  \label{IfromIbar1}\\
  I_-(x,\phi)&=& 
     \displaystyle
     \int_x^0 \frac{dx'}{l|\cos\phi|}\, e^{-(x'-x)/l|\cos\phi|}\,
     \bar I_-(x') + e^{x/l|\cos\phi|}|a(\phi)|^2\,
     \nonumber\\
  &&\qquad\displaystyle
     \times\left( I_0 e^{-L/l|\cos\phi|} + 
       \int_{-L}^0\frac{dx'}{l|\cos\phi|}\, e^{x'/l|\cos\phi|}
       \bar I_+(x') 
     \right)
     ,\qquad\mbox{for $\cos\phi<0$}.
  \label{IfromIbar2}
\end{eqnarray}
\end{mathletters}%
Finally, integration over $\phi$ leads to two coupled integral equations
for the average intensities,%
\begin{mathletters}%
\label{Milne}%
\begin{eqnarray}
  \label{Milne1}
  \bar I_+(x) &=& \int_{-L}^0 dx'\, M_1(x,x')\, \bar I_+(x') +
                  \int_{-L}^0 dx'\, M_2(x,x')\, \bar I_-(x') +
                  Q_1(x)I_0\,,\\
  \label{Milne2}
  \bar I_-(x) &=& \int_{-L}^0 dx'\, M_1(x,x')\, \bar I_-(x') +
                  \int_{-L}^0 dx'\, M_2(x,x')\, \bar I_+(x') +
                  Q_2(x)I_0\,.
\end{eqnarray}
\end{mathletters}%
We have defined the following kernels and source terms:
\begin{mathletters}
\begin{eqnarray}
  \label{Milne3}
  M_1(x,x')     &=& \frac1\pi \int_0^{\pi/2} \frac {d\phi}{l\cos\phi}\,
                  e^{-|x-x'|/l\cos\phi}
      = \frac1{\pi l} K_0(|x-x'|/l),\\
  \label{Milne4}
  M_2(x,x')&=& \frac1\pi \int_0^{\pi/2}\frac {d\phi}{l\cos\phi}\,
                  e^{( x +x') /l\cos\phi}\; |a(\phi)|^2,\\
  \label{Milne5}
  Q_1(x)&=&\frac1\pi \int_0^{\pi/2} d\phi\,e^{-(L+x)/l\cos\phi},\\
  \label{Milne6}
  Q_2(x)    &=& \frac1\pi \int_0^{\pi/2} d\phi\, 
                  e^{-(L-x)/l\cos\phi}\;|a(\phi)|^2,
\end{eqnarray}
\end{mathletters}%
where $K_0$ is a Bessel function.

\begin{figure}
\centerline{\psfig{file=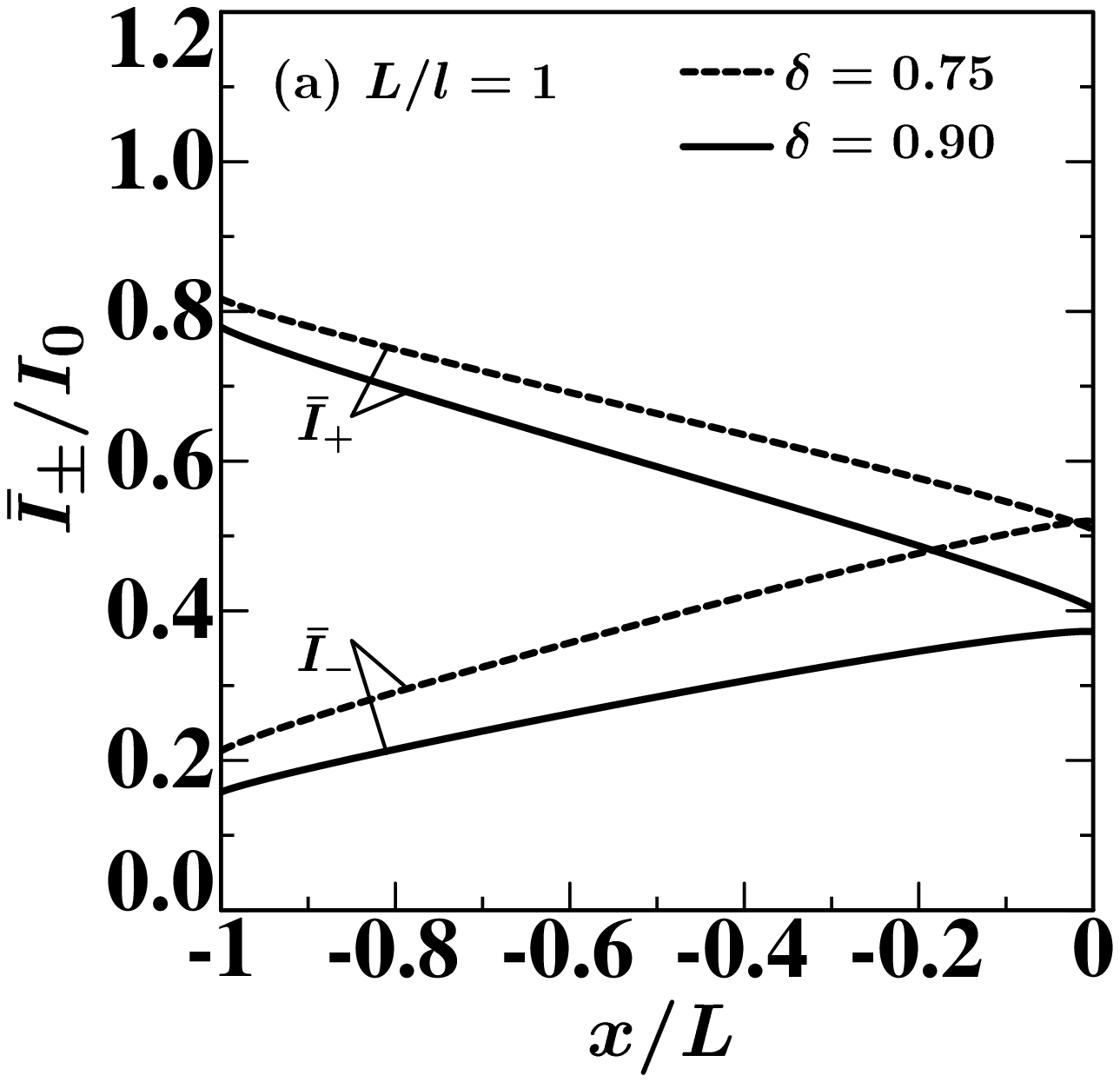,width=0.5\hsize}}
\centerline{\psfig{file=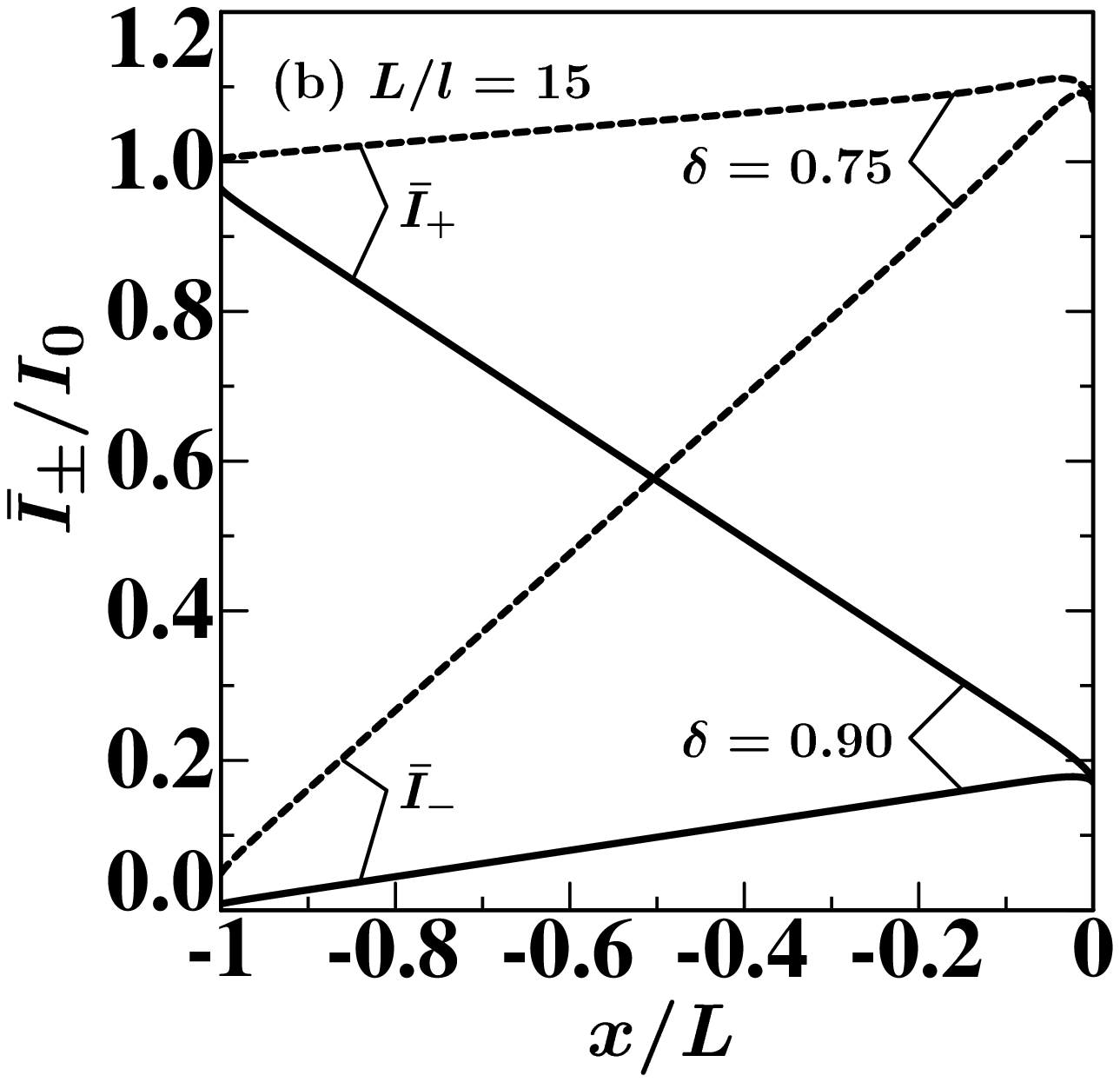,width=0.5\hsize}}
\medskip
\caption{Intensity profiles in the disordered medium, computed from the
exact
numerical solution of the Boltzmann equation,
for $\alpha=\pi/4$ and two values of $\delta$.
(a) is for a nearly ballistic system
($L/l=1$), (b) is for a diffusive system ($L/l=15$).}
\label{profiles}
\end{figure}

Eq.~(\ref{Milne}) is the analogue for the present problem involving two
coupled frequencies of the Schwarzschild-Milne equation in the theory of
radiative transfer\cite{Chan,Ish77}. We have solved it numerically by
discretizing with respect to $x$ so that the integral equation becomes a
matrix equation. From the average intensities
$\bar I_\pm(x)$ one finds the intensities $I_\pm(x,\phi)$ using
Eqs.~(\ref{Formalsol1}) and~(\ref{Formalsol2}). The reflectances~$R_\pm$ 
then follow
{}from Eqs.~(\ref{jdef})--(\ref{Rplusdef}).
For numerical stability we have imposed a cut-off on the rapidly
oscillating function $a(\phi)$ at grazing incidence, by setting
$a(\phi) =0$ for $0.497\,\pi<\phi<\case12\pi$.

\begin{figure}
\centerline{\psfig{file=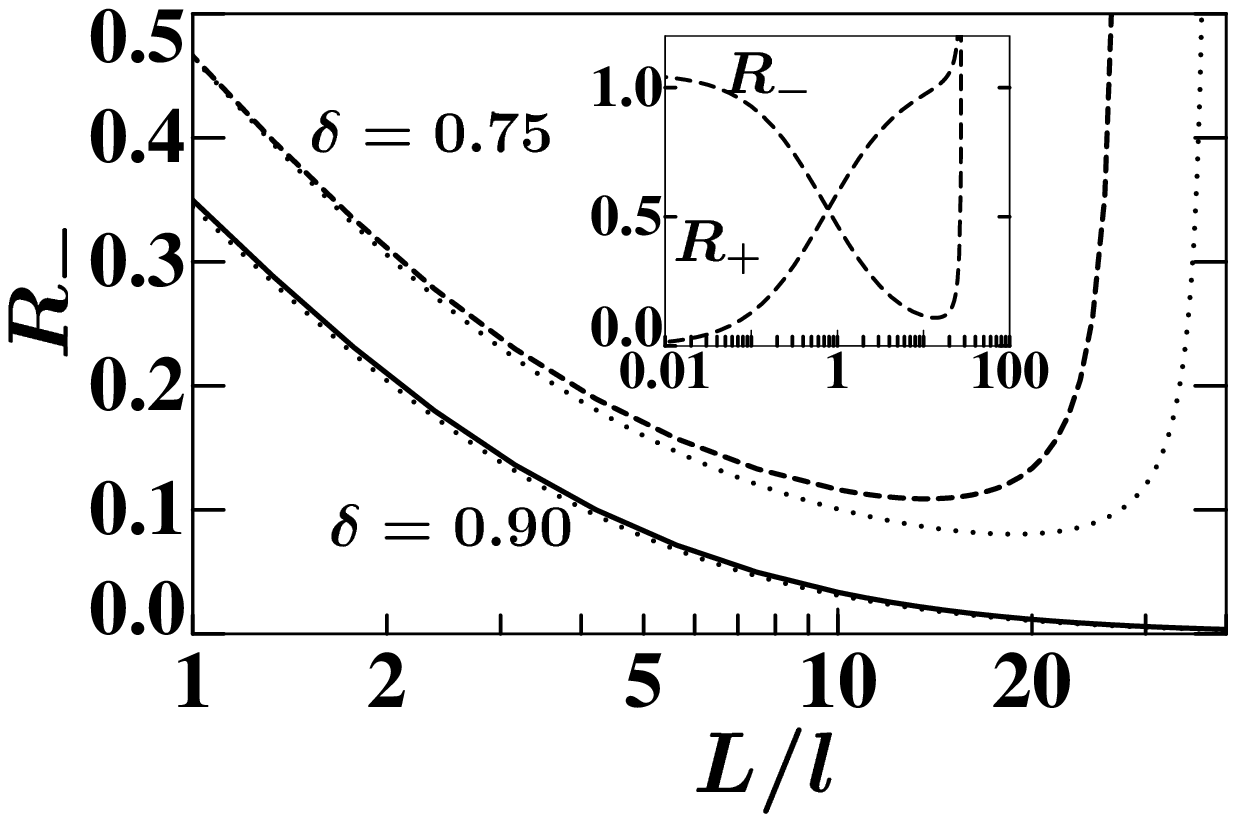,width=0.5\hsize}}
\medskip
\caption{Reflectance $R_-$ as function of $L/l$, computed
{}from
the exact solution of the Boltzmann equation for $\alpha=\pi/4$ and
$\delta=0.75$ (dashed curve), $\delta=0.90$ (solid curve).
The dotted curves are the
approximate result~(\protect\ref{ballmin}), in which angular
correlations are neglected. The inset shows the exact reflectances
$R_\pm$ for $\delta=0.75$, over a broader range of $L/l$ (logarithmic
scale).
For $\delta=0.75$ the reflectances diverge at $L/l=28$. No divergence
occurs for $\delta=0.90$.}
\label{reflections}
\end{figure}

\begin{figure}
\centerline{\psfig{file=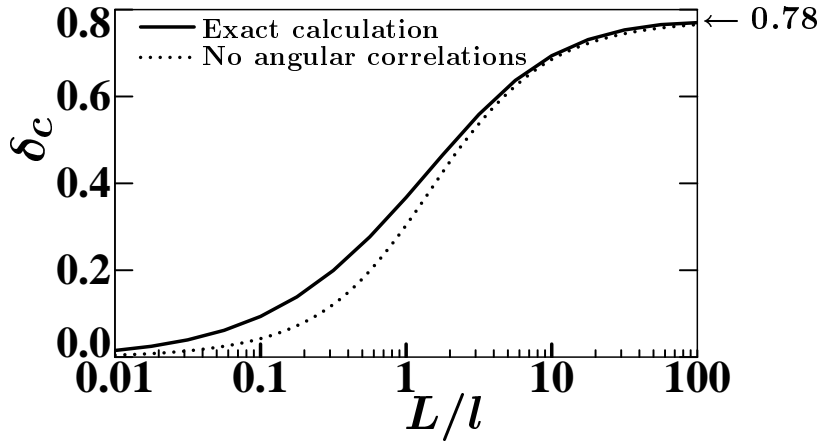,width=0.65\hsize}}
\medskip
\caption{A stationary solution to the Boltzmann equation requires
$\delta>\delta_c$. The solid curve is the exact result for $\delta_c$
(at fixed $\alpha=\pi/4$, as function of $L/l$), the dotted curve
follows from Eq.~(\protect\ref{ballminplus}), obtained by neglecting
angular
correlations.}
\label{singularity}
\end{figure}

In Figs.~\ref{profiles} and~\ref{reflections} we show results for $\bar
I_\pm(x)$ and $R_\pm$ for $\alpha=\pi/4$ and $\delta=0.75$ and $0.9$.
For $\delta=0.75$ there is an effective gain at the mirror ($A>1$), while for
$\delta=0.9$ there is an effective loss ($A<1$). 
For an ordinary mirror one can show that $\bar I_\pm(0)=\case12I_0$. 
Instead, we find that $\bar I_-(0)>\bar I_+(0) >\case12I_0$ for
$\delta=0.75$, indicating gain, and $\bar I_-(0)<\bar I_+(0)<\case12I_0$ for
$\delta=0.9$, indicating loss.
In each case the density profiles are approximately linear in the bulk,
with some bending near the boundaries at $x=-L$ and $x=0$.
For $\delta=0.75$, both
$R_-$ and $R_+$ diverge when $L/l=28$, while for $\delta=0.9$ no
such divergence occurs. As discussed earlier, the divergence indicates
that for $\delta=0.75$ and $L/l> 28$ there is no stationary
solution to the Boltzmann equation.  For fixed $L/l$ and $\alpha$,
the divergence of $R_\pm$ occurs at a critical value~$\delta_c$, 
such that a stationary solution requires $\delta>\delta_c$.
The dependence of $\delta_c$ on $L/l$ at fixed $\alpha=\pi/4$ is plotted in
Fig.~\ref{singularity}.

In Figs.~\ref{reflections} and~\ref{singularity} we also compare the exact
numerical solution of the Boltzmann equation of this subsection
with the approximate analytical solution~(\ref{ballminplus}) of the previous
subsection.  As one can see, the
agreement with the exact results is quite good. 

\section{Phase-coherent solution}
\label{SecPhase}
\subsection{Scattering matrices}
\label{SSecScat}
We now turn to a phase-coherent description of the scattering problem.
To define finite-dimensional scattering matrices we embed the disordered
medium in a waveguide (width~$W$), containing $N_\pm = {\rm Int}
(\omega_\pm W/c\pi) \gg 1$ propagating modes at frequency $\omega_\pm$.
A basis of scattering states consists of the complex fields
\begin{mathletters}
  \label{basisdef}
\begin{eqnarray}
  E_{\pm,n}^>(x,y,t) &=& k_{\pm,n}^{-1/2}\, \sin\left(\frac{n\pi y}{W}\right)\;
       \exp(ik_{\pm,n} x-i\omega_\pm t),\\
  E_{\pm,n}^<(x,y,t) &=& k_{\pm,n}^{-1/2}\, \sin\left(\frac{n\pi y}{W}\right)\;
       \exp(-ik_{\pm,n} x-i\omega_\pm t).
\end{eqnarray}
\end{mathletters}%
Here $n=1,2,\ldots,N_\pm$ is the mode index and the superscript $>$
($<$) indicates a wave moving to the right (left), with frequency
$\omega_\pm = \omega_0\pm\Delta\omega$ and wavenumber
\begin{equation}
  k_{\pm,n} = (\omega_\pm^2/c^2 - n^2\pi^2/W^2)^{1/2}.
  \label{wavenumdef}
\end{equation}
The normalization in Eq.~(\ref{basisdef}) has been chosen such that each
wave carries the same flux.

With respect to the basis~(\ref{basisdef}), incoming and outgoing waves
are decomposed as
\begin{mathletters}
\begin{eqnarray}
  E^{\rm in} &=& \sum_{n=1}^{N_+} u_{+,n} E_{+,n}^> +
                 \sum_{n=1}^{N_-} u_{-,n} E_{-,n}^>, \\
  E^{\rm out}&=& \sum_{n=1}^{N_+} v_{+,n} E_{+,n}^< +
                 \sum_{n=1}^{N_-} v_{-,n} E_{-,n}^< .
\end{eqnarray}
\end{mathletters}%
The complex coefficients are combined into two vectors
\begin{mathletters}
\label{uvdef}
\begin{eqnarray}
  {\bf u} &=& (u  _{+,1}, u  _{+,2},\ldots, u  _{+,N_+},
         u^*_{-,1}, u^*_{-,2},\ldots, u^*_{-,N_-})^{\rm T},\\
  {\bf v} &=& (v  _{+,1}, v  _{+,2},\ldots, v  _{+,N_+},
         v^*_{-,1}, v^*_{-,2},\ldots, v^*_{-,N_-})^{\rm T}.
\end{eqnarray}
\end{mathletters}%
The reflection matrix ${\bf r}$ relates ${\bf u}$ to ${\bf v}$,
\begin{equation}
  {\bf v} = {\bf r}{\bf u},\qquad {\bf r}=
  \pmatrix{{\bf r}_{++} & {\bf r}_{+-}\cr {\bf r}_{-+}&{\bf r}_{--}}.
  \label{Sdef}
\end{equation}
The dimension of ${\bf r}$ is $(N_++N_-)\times(N_++N_-)$, the submatrices
${\bf r}_{\pm,\pm}$ have dimensions $N_\pm\times N_\pm$. For
$\Delta\omega\ll\omega_0$ we may neglect the difference between $N_+$
and $N_-$ and replace both by $N={\rm Int} (k_0 W/\pi)$.

In the absence of disorder the reflection matrix is entirely determined
by the phase-conjugating mirror,
\begin{mathletters}
\label{adef2}
\begin{eqnarray}
  {\bf r}_{\rm PCM} &=& \pmatrix{0&-i{\bf a} e^{-i\psi}\cr
                           i{\bf a} e^{i\psi}&0},\\
  a_{mn} &=& a(\phi_n)\delta_{mn},\qquad\phi_n = \arcsin(n\pi/k_0W).
\end{eqnarray}
\end{mathletters}%
The elements of the $N\times N$ diagonal matrix ${\bf a}$ are obtained
{}from Eq.~(\ref{adeffull}) upon substitution of $\phi$ by $\phi_n$, being the
angle of incidence associated with mode $n$. (The difference in angle
between the two frequencies $\omega_+$ and $\omega_-$ can be neglected
if $\Delta\omega\ll \omega_0$.) The angular average~(\ref{Adef}) of the
reflectance corresponds to the modal average
\begin{equation}
  A = \frac1N{\rm Tr}\  {\bf a}{\bf a}^\dagger.
  \label{Adef2}
\end{equation}
In the limit $N\to\infty$ the two averages are identical.

The disordered medium in front of the phase-conjugating mirror does not
couple $\omega_+$ and $\omega_-$. Its scattering properties at frequency
$\omega$ are described by two $N\times N$ transmission matrices
${\bf t}_{21}(\omega)$ and ${\bf t}_{12}(\omega)$ (transmission from left to right
and from right to left) plus two $N\times N$ reflection matrices
${\bf r}_{11}(\omega)$ and ${\bf r}_{22}(\omega)$ (reflection from left to left and
{}from right to right). Taken together, these four matrices constitute a
$2N\times 2N$ scattering matrix 
\begin{equation}
  {\bf S}_{\rm disorder}(\omega) = 
  \pmatrix{ {\bf r}_{11}(\omega) & {\bf t}_{12}(\omega) \cr
            {\bf t}_{21}(\omega) & {\bf r}_{22}(\omega) },
\end{equation}
which is
unitary (because of flux conservation) and symmetric (because of
time-reversal invariance).
It is simple algebra to express the scattering matrix ${\bf r}$ of the entire
system in terms of the scattering matrices ${\bf r}_{\rm PCM}$ and ${\bf
S}_{\rm
disorder}$ of the phase-conjugating mirror and the disordered region
separately. The result is
\begin{mathletters}
\label{Ssubmatdef}
\begin{eqnarray}
  {\bf r}_{++}^{\vphantom{*}} &=& {\bf r}_{11}^{\vphantom{*}}(\omega_+^{\vphantom{*}}) +
                     {\bf t}_{12}^{\vphantom{*}}  (\omega_+^{\vphantom{*}}) {\bf a} 
                     {\bf r}_{22}^*(\omega_-^{\vphantom{*}}) {\bf a}
     [1-{\bf r}_{22}^{\vphantom{*}}(\omega_+^{\vphantom{*}}) {\bf a}
        {\bf r}_{22}^*(\omega_-^{\vphantom{*}}) {\bf a}]^{-1} {\bf t}_{21}^{\vphantom{*}}(\omega_+^{\vphantom{*}}),\\
  {\bf r}_{--}^{\vphantom{*}} &=& {\bf r}_{11}^*(\omega_-^{\vphantom{*}}) +
                     {\bf t}_{12}^*(\omega_-^{\vphantom{*}}) {\bf a}
                     {\bf r}_{22}^{\vphantom{*}}  (\omega_+^{\vphantom{*}}) {\bf a}
     [1-{\bf r}_{22}^*(\omega_-^{\vphantom{*}}) {\bf a}  
        {\bf r}_{22}^{\vphantom{*}}  (\omega_+^{\vphantom{*}}) {\bf a}]^{-1} {\bf t}^*_{21}(\omega_-^{\vphantom{*}}),\\
  {\bf r}_{-+}^{\vphantom{*}} &=& i e^{i\psi} {\bf t}_{12}^*(\omega_-^{\vphantom{*}}) {\bf a}
     [1-{\bf r}_{22}^{\vphantom{*}}(\omega_+^{\vphantom{*}}) {\bf a}  
        {\bf r}_{22}^*(\omega_-^{\vphantom{*}}) {\bf a}]^{-1} {\bf t}_{21}^{\vphantom{*}}(\omega_+^{\vphantom{*}}),\\
  {\bf r}_{+-}^{\vphantom{*}} &=& -ie^{-i\psi} {\bf t}_{12}^{\vphantom{*}}(\omega_+^{\vphantom{*}}) {\bf a}
     [1-{\bf r}_{22}^*(\omega_-^{\vphantom{*}}) {\bf a}  
        {\bf r}_{22}^{\vphantom{*}}  (\omega_+^{\vphantom{*}}) {\bf a}]^{-1} {\bf t}^*_{21}(\omega_-^{\vphantom{*}}).
\end{eqnarray}
\end{mathletters}%

We seek the reflectances 
\begin{equation}
  R_- = \frac 1{N_+} 
     {\rm Tr}\  {\bf r}_{-+}^{\vphantom{\dagger}} {\bf r}_{-+}^\dagger;
  \qquad\qquad
  R_+ = \frac 1{N_+} 
     {\rm Tr}\  {\bf r}_{++}^{\vphantom{\dagger}} {\bf r}_{++}^\dagger,
  \label{R-R+def}
\end{equation}
averaged over the disorder.
We will do this analytically, using random-matrix theory\cite{Sto91},
and numerically, using the recursive Green
function technique\cite{Bar91}.
We consider two different regimes, depending on the relative magnitude of
$\Delta\omega$ and $1/\tau_{\rm dwell}$, where $\tau_{\rm dwell}\simeq
L^2/cl$ is the mean dwell time of a photon in the disordered medium.
If $\tau_{\rm dwell}\Delta\omega\ll 1$ the difference between ${\bf S}_{\rm
disorder}(\omega_+)$ and  ${\bf S}_{\rm disorder}(\omega_-)$ is insignificant,
because the phase shifts accumulated in a time $\tau_{\rm dwell}$ are
approximately the same for frequencies $\omega_+$ and $\omega_-$. 
We call this the {\em coherent} regime.
If $\tau_{\rm dwell}\Delta\omega\gg1$, on the contrary, phase shifts at
$\omega_+$ and $\omega_-$ are essentially uncorrelated, so that ${\bf
S}_{\rm
disorder}(\omega_+)$ and  ${\bf S}_{\rm disorder}(\omega_-)$  are independent.
We call this the {\em incoherent} regime.

\subsection{Random-matrix theory}
\label{SSecRandom}

Without loss of generality the reflection and transmission matrices of
the disordered region can be decomposed as \cite{Sto91}
\begin{mathletters}
  \label{polar}
\begin{eqnarray}
  {\bf r}_{11}^{\vphantom{\rm T}}(\omega_\pm^{\vphantom{\rm T}}) 
     &\displaystyle = i\,{\bf V}_\pm^{\vphantom{\rm T}}
                      \sqrt{\displaystyle 1-{\bf T}_\pm^{\vphantom{\rm T}}}
                    \;{\bf V}_\pm^{\rm T},
  \qquad\egroup$\hfill$\bgroup &
  {\bf r}_{22}^{\vphantom{\rm T}}(\omega_\pm^{\vphantom{\rm T}}) 
                    = i\,{\bf U}_\pm^{\vphantom{\rm T}}
                      \sqrt{\displaystyle 1-{\bf T}_\pm^{\vphantom{\rm T}}}
                    \;{\bf U}_\pm^{\rm T},\\
  {\bf t}_{12}^{\vphantom{\rm T}}(\omega_\pm^{\vphantom{\rm T}})
      &\displaystyle = {\bf V}_\pm^{\vphantom{\rm T}}
                       \sqrt{\displaystyle {\bf T}_\pm^{\vphantom{\rm T}}}
                     \;{\bf U}_\pm^{\rm T},
  \qquad\egroup$\hfill$\bgroup &
  {\bf t}_{21}^{\vphantom{\rm T}}(\omega_\pm^{\vphantom{\rm T}})
                     = {\bf U}_\pm^{\vphantom{\rm T}}
                       \sqrt{\displaystyle {\bf T}_{\pm}^{\vphantom{\rm T}}}
                       %\sqrt{{\bf T}_\pm^{{\rm T}}}
                     \;{\bf V}_\pm^{\rm T}.
\end{eqnarray}
\end{mathletters}%
Here ${\bf U}_\pm$ and ${\bf V}_\pm$ are  $N\times N$ unitary
matrices (we take $N_+=N_-=N$ in this subsection)
and ${\bf T}_\pm$ is a diagonal matrix with the 
transmission eigenvalues $\tau_{\pm,n}\in[0,1]$ on the diagonal. 
The subscript $\pm$ refers to the two frequencies $\omega_+$ and $\omega_-$. 
In this so-called ``polar decomposition'' the
reflectances $R_\pm$ take the form
\begin{mathletters}
\label{reflect}
\begin{eqnarray}
  R_- &=& \frac 1N {\rm Tr}\  
    {\bf T}_- \mbox{\boldmath$\Omega$}
   \left(1-\sqrt{1-{\bf T}_+}\,\mbox{\boldmath$\Omega$}^{\rm T}
           \sqrt{1-{\bf T}_-}\,\mbox{\boldmath$\Omega$}\right)^{-1}
   \nonumber\\
   && \qquad\qquad\mbox{}\cdot{\bf T}_+
   \left(1-\mbox{\boldmath$\Omega$}^\dagger\sqrt{1-{\bf T}_-}\,
           \mbox{\boldmath$\Omega$}^*      \sqrt{1-{\bf T}_+}\right)^{-1}
   \mbox{\boldmath$\Omega$}^\dagger,
  \label{R-inmat}\\
  R_+ &=& \frac1N{\rm Tr}\ (1-{\bf T}_+) 
      \nonumber\\ &&
   +\;\frac1N{\rm Tr}\  
   {\bf T}_+ \mbox{\boldmath$\Omega$}^{\rm T}\sqrt{1-{\bf T}_-}\, \mbox{\boldmath$\Omega$}
   \left(1-\sqrt{1-{\bf T}_+}\,\mbox{\boldmath$\Omega$}^{\rm T}
           \sqrt{1-{\bf T}_-}\, \mbox{\boldmath$\Omega$}\right)^{-1} 
   \nonumber\\ &&\qquad\qquad
   \mbox{}\cdot{\bf T}_+
   \left(1-\mbox{\boldmath$\Omega$}^\dagger\sqrt{1-{\bf T}_-}\,
           \mbox{\boldmath$\Omega$}^*      \sqrt{1-{\bf T}_+} \right)^{-1} 
   \mbox{\boldmath$\Omega$}^\dagger \sqrt{1-{\bf T}_-}\,\mbox{\boldmath$\Omega$}^*
      \nonumber\\ &&
   -\; \frac1N{\rm Tr}\ {\bf T}_+\sqrt{1-{\bf T}_+}\,
   \left(1-\mbox{\boldmath$\Omega$}^\dagger\sqrt{1-{\bf T}_-}\,
           \mbox{\boldmath$\Omega$}^*      \sqrt{1-{\bf T}_+} \right)^{-1} 
    \mbox{\boldmath$\Omega$}^\dagger \sqrt{1-{\bf T}_-}\,\mbox{\boldmath$\Omega$}^*
      \nonumber\\ &&
   -\; \frac1N{\rm Tr}\ {\bf T}_+\sqrt{1-{\bf T}_+}\, \mbox{\boldmath$\Omega$}^{\rm T}
         \sqrt{1-{\bf T}_-}\, \mbox{\boldmath$\Omega$}
   \left(1-\sqrt{1-{\bf T}_+}\,\mbox{\boldmath$\Omega$}^{\rm T}
           \sqrt{1-{\bf T}_-}\,\mbox{\boldmath$\Omega$}\right)^{-1} ,
   \\
  \mbox{\boldmath$\Omega$} &=& {\bf U}_-^\dagger {\bf a}{\bf 
                             U}_+^{\vphantom{\dagger}}.
\end{eqnarray}
\end{mathletters}%

To compute the averages $\langle R_\pm\rangle$ analytically in the large
$N$-limit we make the isotropy approximation\cite{Sto91}
that the matrices ${\bf
U}_\pm$ and ${\bf V}_\pm$ are uniformly distributed over the unitary
group ${\cal U}(N)$. This approximation corresponds to the neglect of
angular correlations in the radiative-transfer theory
(Sec.~\ref{SSecNeglect}). For $\tau_{\rm dwell}\Delta\omega\ll 1$ we
may identify ${\bf U}_+={\bf U}_-$ and ${\bf V}_+={\bf V}_-$.
For $\tau_{\rm dwell}\Delta\omega\gg 1$ we may assume that ${\bf
U}_+$, ${\bf U}_- $, ${\bf V}_+$, and ${\bf V}_-$ are all
independent. In each case the integration $\int d{\bf U} f({\bf U})$
over ${\cal U}(N)$ with $N\gg1$ can be done using the large
$N$-expansion of Ref.~\cite{Bro96}. The remaining average
over $\tau_{\pm,n}$ can be done using the known 
density $\rho(\tau)$ of the transmission eigenvalues in a disordered
medium\cite{Sto91}.

The calculation is easiest in the incoherent regime 
($\tau_{\rm dwell}\Delta\omega\gg 1$). 
The integration over ${\cal U}(N)$ can be carried out using the 
formula\cite{Bro96}
\begin{eqnarray}
  \int d{\bf U} \int d{\bf V}  \frac1N{\rm Tr}\ 
  ({\bf A}_1 {\bf U} {\bf A}_2 {\bf V} {\bf A}_3 {\bf U} 
           \cdots {\bf A}_p)
  ({\bf B}_1 {\bf U} {\bf B}_2 {\bf V} {\bf B}_3 {\bf U} 
           \cdots {\bf B}_q)^\dagger
  &=&  \delta_{pq} N^{-p} \prod_{i=1}^p
   {\rm Tr}\ {\bf A}_i^{\vphantom{}} {\bf B}_i^\dagger\nonumber\\
  &&\qquad\mbox{}+ {\cal O}(N^{-p-1}).
\end{eqnarray}
To apply this formula we expand the inverse matrices in Eq.~(\ref{reflect})
in a power series in ${\bf U}_\pm$ and integrate term by term over
the independent matrices
${\bf U}_+$ and ${\bf U}_-$. The result is, to leading order in $N$,
\begin{mathletters}
\label{RmpoverU}
\begin{eqnarray}
  \int d{\bf U}_- \int d{\bf U}_+\, R_-
  &=& \sum_{p=0}^\infty T_- T_+ A^{2p+1} 
           (1-T_-)^p (1-T_+)^p\nonumber\\
  &=& \frac{T_- T_+ A}
         {1-(1-T_-)(1-T_+)A^2},\\
  \int d{\bf U}_- \int d{\bf U}_+\, R_+
  &=& 
  %(1-T_+) + \sum_{p=0}^\infty T_+^2 (1-T_-) A^{2p+2} (1-T_-)^p (1-T_+)^p
   1-T_+ + \frac{T_+^2 (1-T_-) A^2}
         {1- (1-T_-)(1-T_+)A^2},
\end{eqnarray}
\end{mathletters}%
where we have defined the modal average
\begin{equation}
  T_\pm = \frac1N{\rm Tr}\  {\bf T}_\pm = \frac1N\sum_{n=1}^N \tau_{\pm,n}.
  \label{Tspecav}
\end{equation}
The modal average $A$ was defined in Eq.~(\ref{Adef2}). The
quantities $T_\pm$ still depend on the configuration of the scatterers,
but the fluctuations around the average $\langle T_\pm\rangle$ are
smaller by an order $1/N$ than the average itself. 
Moreover, the average $\langle T_\pm\rangle$
equals the transmission probability $T$ from radiative transfer theory,
Eq.~(\ref{Tinterpol}), again up to corrections of order $1/N$. Replacing
$T_\pm$ in Eq.~(\ref{RmpoverU}) by $T$ we obtain
\begin{mathletters}
\begin{eqnarray}
  \langle R_-\rangle &=& \frac{T^2 A}{1-(1-T)^2 A^2}, \\
  \langle R_+\rangle &=& 1-T +\frac{T^2(1-T) A^2}{1-(1-T)^2 A^2} ,
\end{eqnarray}
\end{mathletters}%
which is the result~(\ref{ballminplus}) of radiative transfer theory with
neglect of angular correlations. The conclusion is that in the
incoherent regime phase coherence has no effect on the reflectance
of the system to leading order in $N$.

The situation is entirely different in the coherent regime ($\tau_{\rm
dwell}\Delta\omega\ll 1$). To see the difference it is instructive to
first consider the simplified model that the matrix $a_{mn} =
a_0\delta_{mn}$ is proportional to the unit matrix (a scalar). Because ${\bf
U}_-={\bf U}_+$ for $\tau_{\rm dwell}\Delta\omega\ll1$, we then have
$\Omega_{mn} = a_0\delta_{mn}$. There is therefore no average over
${\cal U}(N)$ to perform. We only have to average over one set of
transmission eigenvalues $\tau_{+,n}=\tau_{-,n}\equiv  \tau_n$. 
This average amounts to the integrals
\begin{mathletters}
  \label{R-avint}
\begin{eqnarray}
  \langle R_-\rangle &=& \frac1N\int_0^1 d\tau\rho(\tau)
   \frac{|a_0|^2 \tau^2}{|1-a_0^2+a_0^2 \tau|^2},\\ 
  \langle R_+\rangle  
  &=& 1-\frac1N\int_0^1 d\tau\rho(\tau)
   \;\frac{\tau - |a_0|^4 \tau(1-\tau)}{ \left| 1-a_0^2 + a_0^2 \tau\right|^2}.
\end{eqnarray}
\end{mathletters}%
The density $\rho(\tau)$ for $l\lesssim L\ll Nl$ is given by\cite{Sto91}
 \begin{equation}
  \rho(\tau) = \frac N{2(s+1)}\,\frac1{\tau\sqrt{1-\tau}} + 
     {\cal O} (s+1)^{-4}, \qquad s=\frac{2L}{\pi l}.
  \label{rhodef}
\end{equation}
The density has a cut-off for exponentially small $\tau$, which is irrelevant for 
$\langle R_\pm\rangle$ if $a_0^2\not=1$. Substitution of
Eq.~(\ref{rhodef}) into Eq.~(\ref{R-avint}) yields the average
reflectances 
\begin{mathletters}
  \label{R-simpl}
\begin{eqnarray}
  \langle R_-\rangle &=&  2T
   {\rm Re}\  \frac{a_0^*(a_0^2-1)}{a_0^2-a_0^{*2}}{\rm artanh}\
 a_0^{\vphantom{}}, \\
  \langle R_+\rangle
   &=& 1 -2 T {\rm Re}\  \frac{a_0^*(a_0^{2}-1)} {a_0^2-a_0^{*2}} {\rm
artanh}\  a_0^*,
\end{eqnarray}
\end{mathletters}%
where $T$ is again the transmission probability~(\ref{Tinterpol}) 
{}from radiative transfer theory.
Both quantities have a smooth $L$-dependence, with $\langle R_-\rangle$
decreasing monotonically $\propto1/L$. 
In contrast, radiative transfer theory predicts a non-monotonic
$L$-dependence for $A>1$, leading to a divergence at some $L$. For $A<1$,
radiative transfer theory predicts a quadratic decrease of $\langle R_-\rangle
\propto1/L^2$, for large $L$.
The conclusion is that, in the coherent regime, phase
coherence modifies the reflectance of the phase-conjugating mirror to
leading order in $N$.

\begin{figure}
\centerline{\psfig{file=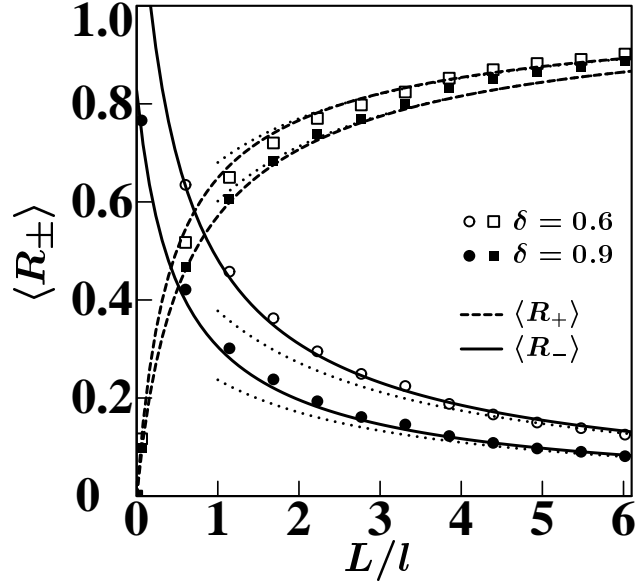,width=0.5\hsize}}
\medskip
\caption{Average reflectances $\langle R_\pm\rangle$
as function of $L/l$ for $\alpha=\pi/4$ and $\delta=0.6$, $0.9$, in the
coherent regime.
The full curves are the analytical results for $N\gg 1$, computed from
Eq.~(\protect\ref{R-R+final}).
The dotted curves are the large $L/l$ limit given by
Eqs.~(\protect\ref{R-simpl}) and~(\protect\ref{a0implicitint}).
Data points are results from numerical simulations.}
\label{figcoherent}
\end{figure}

The result~(\ref{R-simpl}) was obtained for the simplified model of a
scalar reflection matrix ${\bf a}$. The true ${\bf a}$ in
Eq.~(\ref{adef2}) is diagonal but not a scalar. This complicates the
calculation because $\mbox{\boldmath$\Omega$} = {\bf U}_-^\dagger{\bf
a}{\bf U}_+$ then needs to be averaged over ${\cal U}(N)$ even though
${\bf U}_-={\bf U}_+$. The calculation is outlined in
App.~\ref{AppFull}. 
The complete result is a complicated function of $L/l$ (plotted in
Fig.~\ref{figcoherent}).
For $L/l\gg1$ the result takes the form of Eq.~(\ref{R-simpl}), where
now $a_0$ is to be determined from the equation
\begin{equation}
  \frac1N{\rm Tr}\ \frac{{\bf a}}{1-a_0\,{\bf a}} = \frac{a_0}{1-a_0^2}.
  \label{a0implicit}
\end{equation}
In the limit $N\to\infty$ this becomes an integral equation for $a_0$,
\begin{equation}
  \int_0^{\pi/2} d\phi\,\frac{\cos\phi\; a(\phi)}{1-a_0\,a(\phi)} =
       \frac{a_0}{1-a_0^2},
  \label{a0implicitint}
\end{equation}
where $a(\phi)$ is given by Eq.~(\ref{adeffull}).
As shown in Fig.~\ref{figcoherent}, the large-$L$ asymptote
(\ref{R-simpl}), (\ref{a0implicit}) is close to the complete result for
$L\gtrsim l$. In the limit $\delta\to0$ the solution to
Eq.~(\ref{a0implicitint}) is given by $a_0 = 1.284 - 0.0133\,i$,
for $\alpha = \pi/4$.
The corresponding reflectances (for $L\gtrsim l$) are 
$\langle R_-\rangle = 61.1\,l/L$, and
$\langle R_+\rangle = 1+57.7\,l/L$.

To make contact with the work on wave-front reconstruction
\cite{McMichael95,Gu94}, we consider also the case of plane wave ---
rather than diffusive --- illumination. A plane wave incident at frequency
$\omega_{+}$ in mode $n$ is reflected into modes $m=1,2,\ldots N$ at
frequency $\omega_{\pm}$ with probability $\langle|({\bf
r}_{\pm+})_{mn}|^{2}\rangle$. The calculation of this probability
proceeds similarly as the calculation of $R_{-}$. (See
Ref.\ \cite{Melsen95} for the analogous calculation in the case of
Andreev reflection.) 
Using Eqs.~(\ref{Ssubmatdef})--(\ref{reflect}) we can write
\begin{mathletters}
\begin{eqnarray}
  {\bf r}_{-+} &=& ie^{i\psi} {\bf V}_-^* {\bf O} {\bf V}_+^{\rm T},\\
  {\bf O} &=& \sqrt{{\bf T_-}} \mbox{\boldmath$\Omega$}
        (1- \sqrt{1-{\bf T_+}}\mbox{\boldmath$\Omega$}^{\rm T}
            \sqrt{1-{\bf T_-}}\mbox{\boldmath$\Omega$})^{-1}
             \sqrt{{\bf T_+}}.
\end{eqnarray}
\end{mathletters}%
%Hence the value of single reflection amplitude is given by
%\begin{equation}
%  |({\bf r}_{-+})_{mn}|^{2} = \sum_{ijkl} 
%      ({\bf V}_{-})_{ni}^* {\bf O}_{ij}   ({\bf V}_{+})_{mj}
%      ({\bf V}_{-})_{nk}   {\bf O}_{kl}^* ({\bf V}_{+})_{ml}^*.
%\end{equation}
For the coherent regime, we may again identify ${\bf V}_+={\bf V}_-=
{\bf V}$. The
integration over ${\cal U}(N)$ can be performed using\cite{Bro96}
\begin{equation}
  \int d{\bf V}\; V_{nk}^{\vphantom{*}}V_{mj}^{\vphantom{*}} V_{ni}^* V_{ml}^* =
  \frac1{N^2-1}(\delta_{ik}\delta_{jl} + \delta_{mn}\delta_{kl}\delta_{ji})
  -\frac1{N^3-N}(\delta_{kl}\delta_{ji} + \delta_{mn}\delta_{ik}\delta_{jl}).
\end{equation}
We then find
\begin{equation}
   \int d{\bf V}\, |({\bf r}_{-+})_{mn}|^{2} 
     = \frac{1+\delta_{mn}}{N+1} R_-
     + \frac{N\delta_{mn}-1}{N^3-N}\sum_{i\neq j} {\bf O}_{ii}^{\vphantom{*}}
         {\bf O}_{jj}^*.
\end{equation}
In the limit of large $N$ we can write $\sum_{i\neq j} 
{\bf O}_{ii}^{\vphantom{*}}{\bf O}_{jj}^* = |{\rm Tr}\
{\bf O}|^2$. In the same way as
before, for $L\gg l$, this trace can be expressed in terms of $a_0$, 
where $a_0$ can be found  form Eq.~(\ref{adef2}): $N^{-1}  {\rm Tr}\ 
{\bf O} = T\,{\rm artanh}\  a_0$. The result for the averages is
then
\begin{mathletters}
  \label{rmn}
\begin{eqnarray}
   \langle|({\bf r}_{-+})_{nn}|^{2}\rangle &=&
          T^{2}|{\rm artanh}\,a_{0}|^{2},\\
  \langle|({\bf r}_{-+})_{mn}|^{2}\rangle &=&
      N^{-1}\langle R_{-}\rangle,\qquad m\neq n.
\end{eqnarray}
\end{mathletters}%
The incident plane wave is reconstructed with an intensity $\propto
T^{2}$, in agreement with Refs.~\onlinecite{McMichael95,Gu94}.
In the coherent regime, off-diagonal ($m\neq n$) and diagonal ($m=n$) reflection
probabilities differ by a large factor of order $NT$. 

In the incoherent regime, the matrices ${\bf V}_+$ and ${\bf V}_-$ are
independent. Integration over ${\cal U}(N)$ results in integrals of the
form $\int d{\bf V}\; V_{nk}^{\vphantom{*}}  V_{ni}^* = N^{-1} 
\delta_{ik}$. Then the off-diagonal and diagonal  reflection
probabilities are both given by 
\begin{equation}
  \langle|({\bf r}_{-+})_{mn}|^{2}\rangle=N^{-1}\langle R_{-}\rangle,
\end{equation}
so there is no peak in the reflected intensity at the angle of
incidence. This holds for every $N$ and $L$.

For both the  incoherent and the coherent regime we find
for the reflection without frequency shift $(\omega_+\to\omega_+)$ the
probability
\begin{equation}
  \langle|({\bf r}_{++})_{mn}|^{2}\rangle= \frac{1+\delta_{mn}}{1+N}
         \langle R_+\rangle.
\end{equation}
Here we see a much smaller backscattering peak, where the diagonal
reflection probability is only twice as large as the off-diagonal
reflection probability\cite{Mel88a}. This factor is independent of the
phase-conjugating mirror, and exists entirely because of time-reversal
symmetry\cite{Bergmann}.

\subsection{Numerical simulations}
\label{SecSimul}

To test the analytical predictions of random-matrix theory  we
have carried out numerical simulations. The Helmholtz equation,
\begin{equation}
  (-\nabla^2 - \varepsilon\omega_\pm^2/c^2) {\cal E} =0,
\end{equation}
is discretized on a square lattice (lattice constant $d$, 
length $L$, width $W$). 
Disorder is introduced by letting
the relative dielectric constant $\varepsilon$ fluctuate from site to
site between $1\pm\delta\varepsilon$. 
Using the method of recursive Green functions\cite{Bar91} we compute the
scattering matrix ${\bf S}_{\rm disorder}(\omega)$ of the disordered medium at
frequencies $\omega_+$ and $\omega_-$.
The reflection matrix ${\bf r}_{\rm PCM}$ of the phase-conjugating mirror is
calculated by discretizing Eq.~(\ref{BdGlike}). From  ${\bf S}_{\rm
disorder}(\omega_\pm)$ and ${\bf r}_{\rm PCM}$ we obtain the reflection
matrix ${\bf r}$ of the entire system, and hence the 
reflectances~(\ref{R-R+def}).

We took $W=51\,d$, $\delta\varepsilon=0.5$, $\alpha=\pi/4$, and varied
$\delta$ and $L$. For the coherent case we took $\omega_+=\omega_- =
1.252\,c/d$, and for the incoherent case $\omega_+=1.252\,c/d$,
$\omega_-=1.166\,c/d$. These parameters correspond to $N_+=22$,
$l_+=15.5\,d$ at frequency $\omega_+$. The mean free path is determined
using Eq.~(\ref{Tinterpol}), which holds up to small corrections of
order $N^{-1}$. In the incoherent case we have $N_-=20$,
$l_-=20.1\,d$. This leads to $\Delta\omega = 0.043\,c/d$ and a dwell
time for $L/l\simeq 3$ of $\tau_{\rm dwell}\simeq L/cl\simeq150\, d/c$. 
Hence we have
$\tau_{\rm dwell}\Delta\omega\simeq 6.5$, which should be well in the
incoherent regime.
For comparison with random-matrix theory, we take the large $N$-limit
and use the value $l_+$ for $l$.

\begin{figure}
\centerline{\psfig{file=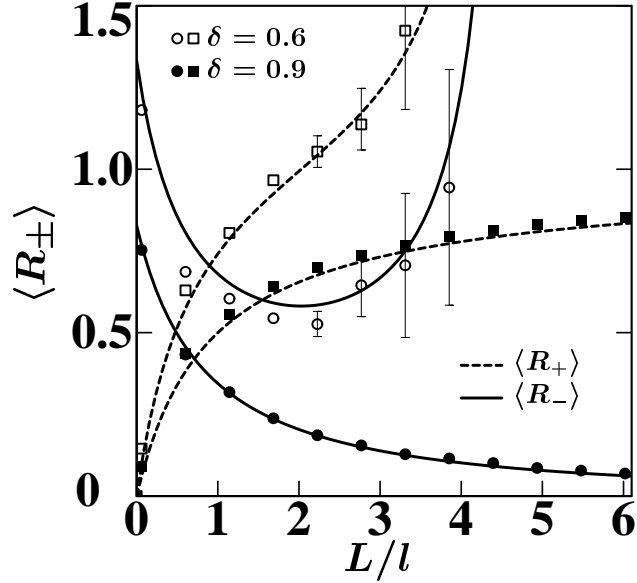,width=0.5\hsize}}
\medskip
\caption{Average  reflectances $\langle R_\pm\rangle$
as function of $L/l$ for $\alpha=\pi/4$ and $\delta=0.6$, $0.9$, in the
incoherent regime.
The curves are the analytical results for $N\gg 1$, computed from
Eq.~(\protect\ref{ballminplus}).
Data points are results from numerical simulations.
(Statistical error bars are shown when they are larger than the size of
the
marker.)}
\label{fignondeg}
\end{figure}

The numerical results are shown in Figs.~\ref{figcoherent} (coherent
regime)
and~\ref{fignondeg} (incoherent regime),
for $\delta = 0.6$ and~$0.9$.
As we can see the agreement with the analytical theory
is quite satisfactory. The rapid rise of $\langle R_\pm\rangle$ in the
incoherent regime for the smallest $\delta$ is accompanied by
large statistical fluctuations, which make an accurate comparison more
difficult. Still, the striking differences between the coherent and
incoherent regimes predicted by the random-matrix theory are confirmed
by the simulations.

\begin{figure}
\centerline{
\psfig{file=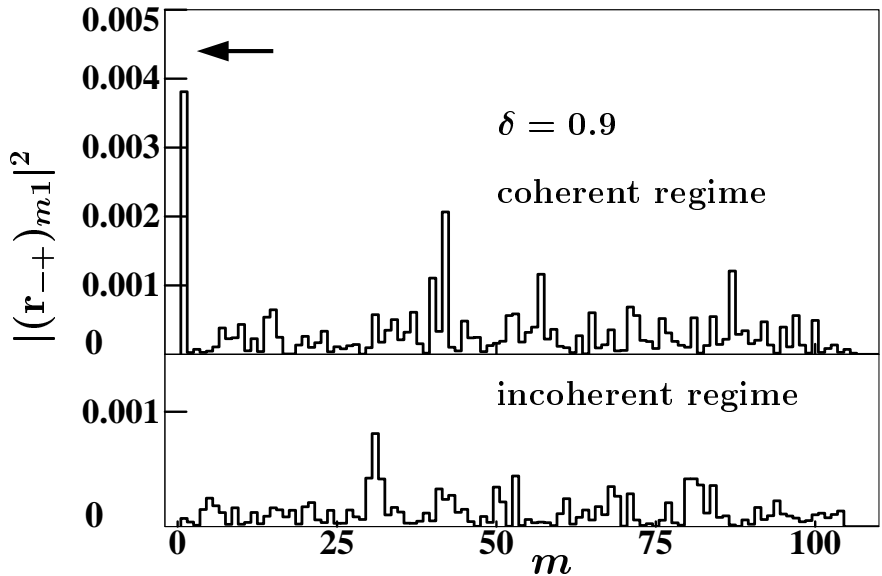,width=0.7\hsize}}
\medskip
\caption{Histograms for the modal distribution $|({\bf
r}_{-+})_{m1}|^2$ of the reflection probability with frequency shift
for normal incidence. The results are for a
single realization of the disorder at $W=L=251\,d$ ($L/l = 16.2$),
$\alpha=\pi/4$, and $\delta=0.9$.
The arrow indicates the theoretical value $\langle|({\bf
r}_{-+})_{11}|^2\rangle$ from Eq.~(\protect\ref{rmn}), representing the
ensemble average in the large $N$-limit.
}
\label{figmodes}
\end{figure}

We have also studied the backscattering peak for plane-wave
illumination. We considered a square sample ($W=L=251\, d$) with
$\alpha=\pi/4$, $\delta=0.9$. We calculated the reflection probabilities
$|(r_{-+})_{mn}|^2$ for normal incidence
($n=1$) in both the coherent and the incoherent regimes. The numerical
results for a single realization of the disorder are
shown in Fig.~\ref{figmodes}. The arrow denotes the analytical ensemble
average~(\ref{rmn}) of the backscattering peak in the large-$N$ limit,
which is consistent with the numerical data. Notice the absence of a 
backscattering peak in the incoherent regime.

\section{Comparison with Andreev reflection}
\label{SecConclusions}

We have studied the reflection of light by a disordered dielectric 
medium in front of a phase-conjugating mirror. This problem 
has an electronic analogue\cite{Lenstra,Hout91}. 
The electronic disordered system consists of a metal, in which 
electron or hole excitations are scattered elastically by
randomly placed impurities. Retro-reflection
at the phase-conjugating mirror is analogous to Andreev reflection at 
the interface with a superconductor. 
The Fermi energy $E_F$ plays the role of the pump frequency $\omega_0$,
while the excitation energy $E$ corresponds to the frequency shift
$\Delta\omega$. 
In spite of these similarities, the optical effects found in this paper
have no electronic analogue. It is instructive to see where the analogy
breaks down.

To this end we compare the wave equation~(\ref{BdGlike}) with the 
Bogoliubov-de Gennes equation\cite{DeGennes}
\begin{equation}
    \pmatrix{H&\Delta\cr\Delta^*&- H}\pmatrix{u\cr v} =
  E \,\pmatrix{u\cr v},
   \label{BdGorig}
\end{equation}
which determines the electron and hole wavefunctions $u$ and $v$.
The Hamiltonian 
\begin{equation}
  H = -\frac{\hbar^2}{2m}\nabla^2 + V - E_F
  \label{Hdef2}
\end{equation}
contains the electrostatic potential $V({\bf r})$, which plays the role
of the dielectric constant. (More precisely, $k_0^2(\varepsilon -1)$
corresponds to $-(2m/\hbar^2) V$.) The role of the nonlinear
susceptibility is played here by the pair  potential $\Delta({\bf r})$, 
which is only nonzero in the superconductor, where it equals
$\Delta_0e^{-i\psi}$.
Comparing Eqs.~(\ref{BdGorig}) and ~(\ref{Hdef2}) for the electronic case
with the optical equations~(\ref{BdGlike}) and~(\ref{Hdef}) 
one notices many similarities, and some differences which amount to a
redefinition of quantities. There is however one essential difference:
the matrix operator in Eq.~(\ref{BdGorig}) is Hermitian,
while that in Eq.~(\ref{BdGlike}) is not, because of an extra minus sign
in one of the off-diagonal elements. This minus sign is the origin
of the difference between Andreev reflection and optical phase
conjugation.

In the optical case a disordered medium becomes transparent ($R_-=1$)
\cite{Wolf,Mittra84} for unit reflectance
at the phase-conjugating mirror ($a=1$). This does
not happen in the electronic case, where $R_-$ is reduced by disorder
even for ideal Andreev reflection.
The reflection matrix of the normal-metal--superconductor (NS) interface,
obtained from Eq.~(\ref{BdGorig}) for $V\equiv 0$, $E\ll \Delta_0\ll
E_F$, is
given by\cite{Andreev64}
\begin{equation}
  {\bf r}_{\rm NS} = \pmatrix{0&-ie^{-i\psi}\cr -ie^{i\psi}&0}.
  \label{rNS}
\end{equation}
Comparison with Eq.~(\ref{adef2}) for ${\bf r}_{\rm PCM}$ shows that
Andreev reflection is independent of the angle of incidence; the matrix
${\bf a}$ in Eq.~(\ref{adef2}) is replaced by the unit matrix in
Eq.~(\ref{rNS}). This is a substantial simplification of the electronic
problem, compared with the optical analogue. The matrix ${\bf r}_{\rm
NS}$
is unitary, in contrast to ${\bf r}_{\rm PCM}$, so that the appearance
of gain or loss at the phase-conjugating mirror has no electronic
counterpart. The reflectance $R_-=1-R_+$ is a monotonically decreasing
function of $L/l$ in the electronic case\cite{LesHouches}, both in the
coherent regime,
\begin{equation}
  R_- = (2+4L/\pi l)^{-1},
  \quad\mbox{if $E\ll \hbar/\tau_{\rm dwell}$ and $L\gtrsim l$,}
  \label{R-deg}
\end{equation}
and in the incoherent regime,
\begin{equation}
  R_- = (1+4L/\pi l)^{-1},
  \quad\mbox{if $E\gg \hbar/\tau_{\rm dwell}$.}
  \label{R-non}
\end{equation}

The result~(\ref{R-non}) is what one obtains from
Eq.~(\ref{ballminplus}) for the case $A=1$ of unit reflectance at the
interface. (The transmittance $T=(1+2L/\pi l)^{-1}$ of the disordered
medium is the same for electrons and photons.) The result~(\ref{R-deg})
however, is {\em not} what one would expect from the optical analogue.
Indeed, Eq.~(\ref{R-avint}) with $a_0=1$ would give $R_-=1$ for all $L$
in the case of unit reflectance at the phase-conjugating mirror. The
reason that the analogy with Andreev reflection breaks down is the
difference of a minus sign in the wave equations~(\ref{BdGlike})
and~(\ref{BdGorig}), which reappears in the reflection
matrices~(\ref{adef2}) and~(\ref{rNS}) for phase conjugation, and
ultimately in the reflectances in the coherent regime:
\begin{mathletters}
\begin{eqnarray}
  R_- &=& \frac1N\,{\rm Tr}\ \left(\frac{{\bf t}{\bf t}^\dagger}
             {1+{\bf r}{\bf r}^\dagger}\right)^2 \not=1
      \quad\mbox{for electrons,}\\
  R_- &=& \frac1N\,{\rm Tr}\ \left(\frac{{\bf t}{\bf t}^\dagger}
             {1-{\bf r}{\bf r}^\dagger}\right)^2 =1
      \quad\mbox{for photons if ${\bf a}\equiv 1$.}
\end{eqnarray}
\end{mathletters}%
Here ${\bf t}$ and ${\bf r}$ are the transmission and reflection
matrices of the disordered medium, which satisfy ${\bf t}{\bf
t}^\dagger + {\bf r}{\bf r}^\dagger =1$.

In conclusion, we have shown that the presence of a phase-conjugating
mirror behind a random medium drastically changes the total reflected
intensity, even when the medium is so disordered that wave-front
reconstruction is ineffective. On increasing the frequency difference
$\Delta\omega$ between the incident radiation and the pump beams, a {\em
minimum\/} in the disorder dependence of the reflected intensity
appears.
In a certain parameter range, the disordered medium reflects less
radiation
on reducing $\Delta\omega$. Experimental observation of this
``darkening''
would be a striking demonstration of phase-shift cancellations in a
random
medium.

\acknowledgements
This work was supported by the ``Stichting voor Fundamenteel Onderzoek
der Materie'' (FOM) and by the ``Nederlandse organisatie voor
Wetenschappelijk Onderzoek'' (NWO).

\appendix
\section{Calculation of the reflectances in the coherent regime}
\label{AppFull}

In Sec.~\ref{SecPhase} we computed the average reflectances 
$\langle R_{\pm} \rangle$ for the
incoherent regime. For the coherent regime we presented only a derivation
for scalar reflection matrix ${\bf a}$. This appendix contains the 
calculation of $\langle
R_{\pm} \rangle$ for arbitrary matrix ${\bf a}$. Our calculation is based on
the diagrammatic method for integration over the unitary group of Refs.\
\onlinecite{ArgamanZee} and \onlinecite{Bro96}. Integrals over the
unitary group are needed for the computation of $\langle R_{\pm} \rangle$
because of the polar decomposition~(\ref{polar}) of the transmission and
reflection matrices. We find it convenient to use a slightly
modified version of the diagrammatic technique, in which we apply the
diagrammatic rules without making explicit use of the polar decomposition. We
first outline the calculation of $\langle R_{\pm} \rangle$ in which the
diagrammatic method is used for the integration of the matrices $U$ and $V$ in
Eq.~(\ref{polar}), and then discuss the modification of the
diagrammatic method.

We start the calculation of $\langle R_{\pm} \rangle$ with the elimination of
the reflection matrix ${\bf r}_{11}(\omega_0)$ and the transmission matrices
${\bf t}_{12}(\omega_0)$ and ${\bf t}_{21}(\omega_0)$ from the reflectances
$R_{+}$ and $R_{-}$ [cf.\ Eqs.~(\ref{Ssubmatdef}) and~(\ref{R-R+def})], in
favour of the matrix ${\bf r} = {\bf r}_{22}(\omega_0)$. The result is
\begin{mathletters} \label{eq:rrdef}
\begin{eqnarray}
  \langle R_{+} \rangle &=& {1 \over N} \mbox{Tr}\, [{\bf s}'({\bf a},{\bf a})
- {\bf s} (1,{\bf a}) - {\bf s} ({\bf a},1) + {\bf s}'(1,1) + {\bf h}(1) +
{\bf h}(1)^{*}], \\
  \langle R_{-} \rangle &=& {1 \over N} \mbox{Tr}\, \{{\bf s} ({\bf a},{\bf a})
- {\bf s}'(1,{\bf a}) - {\bf s}'({\bf a},1) + {\bf s} (1,1) % \nonumber \\ &&
%%\mbox{}
%  -  {\bf a}^{-1*} {\bf h}(1) - {\bf a}^{-1} {\bf h}(1)^{*} + {\bf a}^{-1}
  + {\bf a}^{-1} {\bf a}^{-1*} [ 1 - {\bf h}(1) - {\bf h}(1)^* ]\},
\end{eqnarray}
\end{mathletters}%
where we defined
\begin{mathletters} \label{eq:sshdef}
\begin{eqnarray}
  {\bf s}'({\bf x},{\bf y}) &=& \langle {\bf x} (1 - {\bf r} {\bf a} {\bf r}^*
{\bf a})^{-1} {\bf r} {\bf y}
           {\bf y}^{*} {\bf r}^{*} (1 - {\bf a}^* {\bf r} {\bf a}^* {\bf
r}^*)^{-1} {\bf x}^{*} \rangle, \\
  {\bf s} ({\bf x},{\bf y}) &=& \langle {\bf x} (1 - {\bf r} {\bf a} {\bf r}^*
{\bf a})^{-1} {\bf a}^{-1} {\bf y}
           {\bf y}^{*} {\bf a}^{-1*} (1 - {\bf a}^* {\bf r} {\bf a}^* {\bf
r}^*)^{-1} {\bf x}^{*} \rangle, \\
  {\bf h} ({\bf x})   &=& \langle {\bf x} (1 - {\bf r} {\bf a} {\bf r}^* {\bf
a})^{-1} \rangle.
\end{eqnarray}
\end{mathletters}%
To perform the average over ${\bf r}$, one may use the polar decomposition
[cf. Eq.~(\ref{polar})]
\begin{equation}
  {\bf r} = i{\bf U} \sqrt{1-{\bf T}} {\bf U}^{\rm T} \label{eq:polarr},
\end{equation}
where ${\bf U}$ is a unitary matrix and ${\bf T}$ is the diagonal matrix
containing the $N$ transmission eigenvalues $\tau_{j}$ on the diagonal. The
matrix ${\bf U}$ is a member of the circular unitary ensemble (CUE),
i.e.\ it is uniformly distributed in the unitary group. The 
transmission eigenvalues $\tau_{j}$ have density\cite{Sto91}
\begin{mathletters}
  \label{dens}
\begin{eqnarray}
  \rho(\tau) &=& (2N/\pi)\; {\rm Im}\, U(1/\tau-1-i\,0,s),\\
  U(\zeta,s) &=& {\rm cotanh}\,[\zeta-s U(\zeta,s)],
   \qquad s=2L/\pi l.
\end{eqnarray}
\end{mathletters}%

To integrate the matrix ${\bf U}$ over the unitary group, the matrices
${\bf s}$, $ {\bf s}'$, and ${\bf h}$
are first expanded as a power series in ${\bf U}$. The integration of ${\bf
U}$ is then done using the general expression for the average of a 
polynomial function of ${\bf U}$\cite{Samuel},
\begin{eqnarray} \label{eq:Uperm} \label{eq:Uavg}
% &&
 \left\langle {\bf U}_{a_1 b_1} \ldots {\bf U}_{a_m b_m}
          {\bf U}^{*}_{\alpha_1 \beta_1} \ldots {\bf U}^{*}_{\alpha_n \beta_n}
\right\rangle =
%  \nonumber \\ && \
  \delta_{m,n}
  \sum_{P,P'} V_{c_1,\ldots,c_k} \prod_{j=1}^{n} \delta_{a_j,\alpha_{P(j)}}
\delta_{b_j,\beta_{P'(j)}}.
\end{eqnarray}
Here the summation is over all permutations $P$ and $P'$ of the numbers $1,
\ldots, n$. The numbers $c_1,\ldots,c_k$ denote the {\em cycle structure} of
the permutation $P^{-1} P'$. (The permutation $P^{-1} P'$ can be
uniquely written as a product of disjoint cyclic permutations of lengths $c_1,
\ldots, c_k$, with $n = \sum_{j=1}^{k} c_k$.) To compute $\langle R_{\pm}
\rangle$ in the limit of large $N$, it is sufficient to know the
coefficients $V_{c_1,\ldots,c_k}$ to leading order in $N$. These are given
in Refs.\ \onlinecite{ArgamanZee} and \onlinecite{Bro96}, together
with a diagrammatic method which enables one to restrict the summation over
$P$ and $P'$ to those permutations $P$ and $P'$ of which the contribution to
$\langle R_{\pm} \rangle$ is of maximal order in $N$.

Although the computation of $\langle R_{\pm} \rangle$ is straightforward now,
the actual calculation is rather cumbersome. We find it convenient to
modify the approach of Refs.\
\onlinecite{ArgamanZee} and \onlinecite{Bro96} so that it can be
applied directly to the average over the matrix $r$, without making explicit
use of the polar decomposition (\ref{eq:polarr}). This is possible because the
general structure (\ref{eq:Uperm}) already follows from the invariance of the
distribution of ${\bf U}$ under transformations
\begin{equation}
  {\bf U} \to {\bf V} {\bf U} {\bf V}',
\end{equation}
where ${\bf V}$ is an arbitrary unitary matrix. The fact that ${\bf U}$ itself
is unitary is necessary to compute the value of the coefficients
$V_{c_1,\ldots,c_k}$, but it is not relevant for the general structure
(\ref{eq:Uperm}). Since the matrix ${\bf r}$ is both unitary and symmetric,
its distribution is invariant under transformations
\begin{equation}
  {\bf r} \to {\bf V} {\bf r} {\bf V}^{\rm T} \label{eq:rinvariance}
\end{equation}
that respect the symmetry of ${\bf r}$. The same group of transformations
leaves invariant the circular orthogonal ensemble (COE), 
consisting of
uniformly distributed unitary and symmetric matrices. A diagrammatic technique
for averages over the COE is presented in Ref.\ \onlinecite{Bro96}.
As before, the general structure of the average of a polynomial of a matrix
{}from the COE is entirely determined by the invariance under the
transformations (\ref{eq:rinvariance}), and therefore applies to the
reflection matrix ${\bf r}$ as well. It reads
\cite{Bro96,MelloSeligman}
\begin{equation} \label{eq:Uparamb1} \label{eq:Uavgb1}
  \langle {\bf r}_{a_1 a_2} \ldots {\bf r}_{a_{2n-1} a_{2n}} {\bf
r}^{*}_{\alpha_{1} \alpha_{2}} \ldots {\bf r}^{*}_{\alpha_{2m-1} \alpha_{2m}}
\rangle = \delta_{n,m} \sum_{P} V_{c_1,\ldots,c_k} \prod_{j=1}^{2n}
\delta_{a_j,\alpha_{P(j)}},
\end{equation}
where now the summation is over permutations $P$ of the numbers $1$, \ldots,
$2n$. We may write $P$ as
\begin{equation}
  P = \left(\prod_{j=1}^{n} \sigma_j\right) P_e P_o 
      \left(\prod_{j=1}^{n} \sigma'_j\right),
\end{equation}
where the permutations $\sigma_j$ and $\sigma_j'$ operate on the numbers
$2j-1$ and $2j$, and the permutation $P_e$ ($P_o$) permutes even (odd) numbers
only. The numbers $c_1,\ldots,c_k$ in Eq.\ (\ref{eq:Uavgb1}) are the cycle
structure of the permutation $P_e^{-1} P{\vphantom{-1}}_o$.
The specific values of the coefficients
$V_{c_1,\ldots,c_k}$ for an average of ${\bf r}$ are of course
different from those for the COE.

\begin{figure}
\psfig{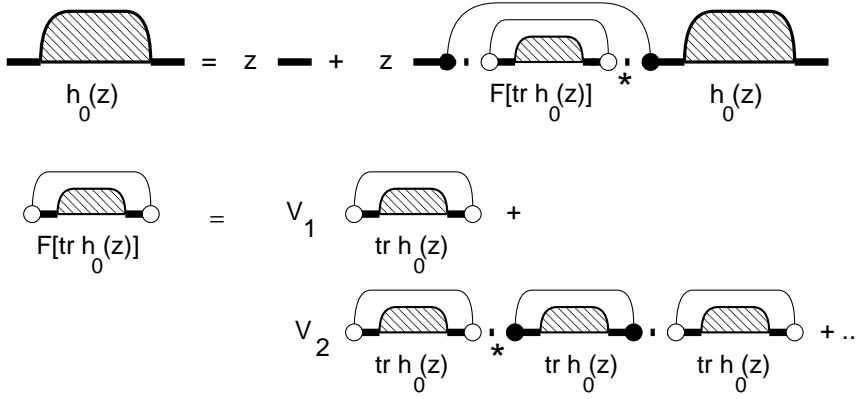}
\caption{Diagrams for the calculation of ${\bf h}_0(z)$.}
\label{fig:diag1}
\end{figure}%

Now that we have identified the formal equivalence of an average over the
(non-unitary) symmetric reflection matrix $r$ and a unitary symmetric matrix
{}from the COE, we can directly apply the diagrammatic rules of Refs.\
\onlinecite{ArgamanZee} and \onlinecite{Bro96} to an average over
the matrix ${\bf r}$, provided we know the coefficients $V_{c_1,\ldots,c_k}$
for the ensemble of reflection matrices $r$ of a disordered waveguide. To find
these coefficients, we use the fact that they
factorize, to leading order in~$N$,
\begin{equation}
  V_{c_1,\ldots,c_k} = \prod_{j=1}^{k} V_{c_k}, \label{eq:Vfact}
\end{equation}
just as they do for the COE. This follows directly from the fact that, to
leading order in $N$, the average $\langle \prod_{j} \mbox{Tr}\, ({\bf r} {\bf
r}^{\dagger})^{c_j} \rangle$ factorizes into $\prod_{j} \langle \mbox{Tr}\,
({\bf r} {\bf r}^{\dagger})^{c_j} \rangle$ \cite{MelloStone}. It remains to
find the coefficients $V_{c}$. Hereto we consider the function
\begin{equation}
  {\bf h}_0(z) = \left\langle {z \over 1 - {\bf r} {\bf r}^{*} z^2}
     \right\rangle.
\end{equation}
We first compute ${\bf h}_0(z)$ from the diagrammatic technique, with a priori
unknown coefficients $V_{c}$. We then compare our result with a calculation of
$\mbox{Tr}\, {\bf h}_0(z)$ from the density of transmission 
eigenvalues~(\ref{dens}). The relevant diagrams
for the diagrammatic calculation are shown in Fig.\ \ref{fig:diag1} (for a
detailed explanation of the diagrammatic notation of Fig.\ \ref{fig:diag1}, we
refer to Ref.\ \onlinecite{Bro96}). The result is a
self-consistency equation for ${\bf h}_0(z)$ that involves the generating
function $F$ of the coefficients $V_c$,
\begin{eqnarray}
  {\bf h}_0(z) &=& {z \openone \over 1 - z F[\mbox{Tr}\, {\bf h}_0(z)]},
\label{eq:h1} \\
  F(x) &=& \sum_{j=c}^{\infty} V_{c}\, x^{2c-1}. \label{eq:Fdef}
\end{eqnarray}
Here $\openone$ is the $N \times N$ unit matrix.
Direct computation of $\mbox{Tr}\, {\bf h}_0(z)$ from the density 
$\rho(\tau)$ of transmission eigenvalues, gives
\begin{equation}
  \mbox{Tr}\, {\bf h}_0(z) = \int_0^1 d\tau\, {\rho(\tau) z \over 1 -
z^2(1-\tau)}. \label{eq:h2}
\end{equation}
Together, Eqs.\ (\ref{eq:Vfact})--(\ref{eq:h2}) determine the coefficients
$V_{c_1,\ldots,c_k}$ needed for the diagrammatic evaluation of $\langle
R_{\pm} \rangle$. In the limit of $L\to\infty$, the density
of transmission eigenvalues tends to $N\delta(\tau)$. Hence ${\bf h}_0(z)
= z/(1-z^2)$ and $F(x) = (\sqrt{N^2+4x^2}-N)/2x$. The corresponding
coefficients $V_{c} = c^{-1} N^{1-2c} {2c-2 \choose c-1}$ are precisely those
of the COE \cite{Bro96}. For finite $L$, the
density $\rho(\tau)$ is no longer a delta function, and hence the
coefficients $V_{c}$ deviate from those of the COE.

\begin{figure}
\psfig{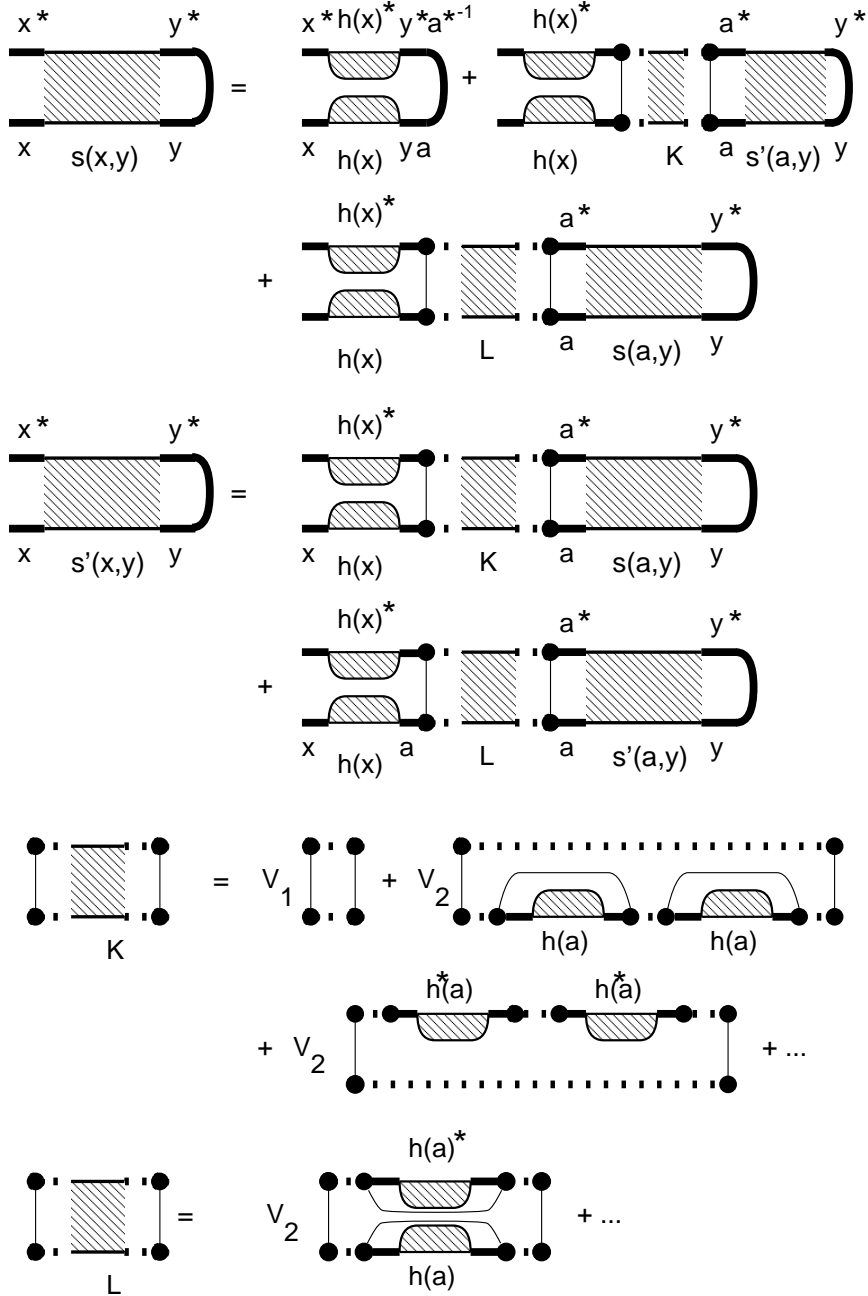}
\caption{Diagrams for the calculation of ${\bf s}({\bf
x},{\bf
 y})$ and ${\bf s}'({\bf x},{\bf y})$.}
\label{fig:diag2}
\end{figure}%

The fact that we can use the diagrammatic rules directly for the average over
$r$ simplifies the calculation considerably. A central role in the calculation
is played by the function ${\bf h}({\bf x})$ defined in Eq.\
(\ref{eq:sshdef}). The diagrams for the calculation of ${\bf h}({\bf x})$ are
similar to those of Fig.\ \ref{fig:diag1}, and the result is a
self-consistency equation for ${\bf h}({\bf x})$
\begin{equation} \label{eq:hsol}
  {\bf h}({\bf x}) = {\bf x} \biglb(1 - {\bf a} F[\mbox{Tr}\, {\bf h}({\bf
a})]\Bigrb)^{-1},
\end{equation}
Notice the formal equivalence with Eq.\ (\ref{eq:h1}). The function $F$ was
defined in Eq.\ (\ref{eq:Fdef}). Using the diagrammatic technique for the
computation of $s$ and $s'$, we find the linear relations
\begin{mathletters} \label{eq:ssol}
\begin{eqnarray}
  {\bf s}({\bf x},{\bf y}) &=& {\bf h}({\bf x}) 
   \left[{\bf a}^{-1} {\bf y} {\bf y}^{*} {\bf a}^{-1*} +
           K\, \mbox{Tr}\, {\bf s'}({\bf a},{\bf y}) + L\, 
        \mbox{Tr}\, {\bf s}({\bf a},{\bf y}) \right]
        {\bf h}({\bf x})^{*}, \\
  {\bf s'}({\bf x},{\bf y}) &=& {\bf h}({\bf x})
    \left[K\, \mbox{Tr}\, {\bf s}({\bf a},{\bf y}) +
    L\, \mbox{Tr}\, {\bf s'}({\bf a},{\bf y}) \right]
     {\bf h}({\bf x})^{*},
\end{eqnarray}
\end{mathletters}%
where we have defined
\begin{mathletters} \label{eq:KLdef}
\begin{eqnarray}
  h &=& \mbox{Tr}\,{\bf h}({\bf a}),\\
  K &=& \sum_{i,j=1}^{\infty} V_{i+j-1} h^{2i-2} (h^*)^{2j-2}
        = \frac{ h F(h) - h^* F(h)^*}{h^2-h^{*2}},\\
  L &=& \sum_{i,j=1}^{\infty} V_{i+j} h^{2i-1} (h^*)^{2j-1}
        = \frac{ h^* F(h) - h F(h)^*}{h^2-h^{*2}}.
\end{eqnarray}
\end{mathletters}%
The relevant diagrams leading to Eqs.\ (\ref{eq:ssol}) and (\ref{eq:KLdef})
are shown in Fig.\ \ref{fig:diag2}. They are similar to those of Ref.\
\onlinecite{ArgamanZee}, where the case of a chaotic cavity was considered,
instead of a disordered waveguide.
Together, Eqs.\ (\ref{eq:hsol})--(\ref{eq:KLdef}) form a closed set of
equations, from which ${\bf s}({\bf x},{\bf y})$, ${\bf s}'({\bf x},{\bf y})$,
and ${\bf h}({\bf x})$ can be calculated. The average reflectances $\langle
R_{\pm} \rangle$ are obtained upon substitution of ${\bf s}({\bf x},{\bf y})$,
${\bf s}'({\bf x},{\bf y})$, and ${\bf h}({\bf x})$ into Eq.\
(\ref{eq:rrdef}).
The final result is expressed as a function of $h=\mbox{Tr}\, {\bf h}({\bf
a})$,
\begin{mathletters}
  \label{R-R+final}
\begin{eqnarray}
  N \langle R_{+} \rangle &=& {(I^2 + J^2) K - 2 J (1 - I L) \over (1 - I L)^2
- (K I)^2} + 2 \mbox{Re}\, \mbox{Tr}\, [1 - {\bf a} F(h)]^{-1},\\
  N \langle R_{-} \rangle &=& {(I + L J^2)(1- I L) + K I J (K I - 2) \over (1
- I L)^2 - (K I)^2} + J |F(h)|^2,
\end{eqnarray}
\end{mathletters}
where we defined
\begin{mathletters}
\begin{eqnarray}
  I &=& \mbox{Tr}\, {\bf a} (1 - {\bf a} F(h))^{-1} (1 - {\bf a}^{*}
F(h)^{*}])^{-1} {\bf a}^{*}, \\
  J &=& \mbox{Tr}\, (1 - {\bf a} F(h))^{-1} (1 - {\bf a}^{*} F(h))^{-1}.
\end{eqnarray}
\end{mathletters}%

These expressions simplify in the large $L/l$-limit, when $\rho(\tau)$
takes the form~(\ref{rhodef}). Substitution in Eq.~(\ref{eq:h2})
gives
\begin{equation}
  {\rm Tr}\  {\bf h}_0(z) = \frac{Nz}{1-z^2} \left(1-
         \frac{z\; {\rm artanh}\,z}{1+s}\right)
     + {\cal O}(1+s)^2,
     \qquad s=2L/\pi l.
\end{equation}
and hence allows us to find $F(z)$ from Eq.~(\ref{eq:h1}).
Expanding the expressions~(\ref{R-R+final})
for $\langle R_\pm\rangle$ and the self-consistency
equation~(\ref{eq:hsol}) to lowest order in $(1+s)^{-1}$ we find the
results~(\ref{R-simpl}) and~(\ref{a0implicit}), with the effective
reflectance $a_0=z$.


\begin{references}

\bibitem{Andreev64} A.~F. Andreev, Zh. Eksp. Teor. Fiz.\ {\bf 46}, 1823
(1964);
{\bf 49}, 655 (1965) [Sov. Phys. JETP {\bf 19}, 1228 (1964); {\bf 22}, 455
(1966)].

\bibitem{Woerdman70} J. P. Woerdman, Optics Comm. {\bf 2}, 212 (1970).

\bibitem{Stepanov71} B. I. Stepanov, E. V. Ivakin, and A. S. Rubanov,
Dokl. Akad. Nauk USSR {\bf 196}, 567 (1971) [Sov. Phys. Dokl. {\bf
16}, 46 (1971)].

\bibitem{Abrikosov} A. A. Abrikosov, {\it Fundamentals of the Theory of
Metals} (North-Holland, Amsterdam, 1988).

\bibitem{Fisher} R.~A. Fisher, editor, {\it Optical Phase Conjugation}
(Academic, New York, 1983).

\bibitem{Zeldovich} B.~Ya. Zel'dovich, N.~F. Pilipetski\u\i, and V.~V.
Shkunov, {\it Principles of Phase Conjugation} (Springer, Berlin,
1985).

\bibitem{Pep86} D.~M. Pepper, Sci. Am. {\bf 254} (1), 56 (1986).

\bibitem{LesHouches} C.~W.~J. Beenakker, in: {\it Mesoscopic Quantum Physics},
edited
by E. Akkermans, G. Montambaux, J.-L. Pichard, and J. Zinn-Justin
(North-Holland, Amsterdam, 1995).


\bibitem{Wolf} Yu. N. Barabanenkov, Yu. A. Kravtsov, V. D. Ozrin, and A.
I. Saichev, in: {\it Progress in Optics XXIX}, edited by E. Wolf
(North-Holland, Amsterdam, 1991).

\bibitem{Mittra84} R. Mittra and T. M. Habashy, J. Opt. Soc. Am. A {\bf
1}, 1103 (1984).

\bibitem{McMichael95} I. McMichael, M. D. Ewbank, and F. Vachss, Optics
Comm. {\bf 119}, 13 (1995).

\bibitem{Gu94} C. Gu and P. Yeh, Optics Comm. {\bf 107}, 353 (1994).

\bibitem{Chan} S. Chandrasekhar, {\it Radiative Transfer} (Dover,
New York, 1960).

\bibitem{Ish77} A. Ishimaru, {\it Wave Propagation and Scattering in
Random Media} (Academic, New York, 1978).

\bibitem{Paa96} J.~C.~J.~Paasschens, P.~W.~Brouwer, and C.~W.~J.~Beenakker,
Europhys.~Lett. (to be published).

\bibitem{Freund88} I. Freund, M. Rosenbluh, R. Berkovits, and M. Kaveh,
Phys. Rev. Lett.\ {\bf 61}, 1214 (1988); M.~I. Mishchenko, J.~M.
Dlugach,
and E.~G. Yanovitskij, J. Quant. Spectrosc. Radiat. Transfer {\bf 47},
401 (1992).

\bibitem{Lenstra} D. Lenstra, in: {\it Huygens Principle 1690--1990;
Theory
and Applications}, edited by H. Blok, H.~A. Ferweda,
and H.~K. Kuiken (North-Holland, Amsterdam 1990).

\bibitem{Hout91} H. van Houten and C.~W.~J. Beenakker, Physica B\ {\bf
175}, 187 (1991).

\bibitem{Yariv77} A. Yariv and D.~M. Pepper, Opt. Lett.\ {\bf 1}, 16
(1977).

\bibitem{Arnold89} H.~F. Arnoldus and T.~F. George, J. Mod. Opt.\ {\bf
36}, 31 (1989).

\bibitem{Jong94} M.~J.~M. de Jong, Phys. Rev. B\ {\bf 49}, 7778 (1994).

\bibitem{Sto91} Two reviews of the random-matrix theory of
phase-coherent scattering are: A.~D. Stone, P.~A. Mello, K.~A. Muttalib,
and
J.-L. Pichard, in: {\it Mesoscopic Phenomena in Solids}, edited by
B.~L. Altshuler, P.~A. Lee, and R.~A. Webb (North-Holland, Amsterdam,
1991);
C.~W.~J. Beenakker, Rev. Mod. Phys., (July, 1997).

\bibitem{Bar91} H.~U.~Baranger, D.~P.~DiVincenzo, R.~A.~Jalabert, and
A.~D.~Stone, Phys.\ Rev.\ B {\bf 44}, 10637 (1991).

\bibitem{Bro96} P.~W. Brouwer and C.~W.~J. Beenakker, J.~ Math.~Phys.
{\bf 37}, 4904 (1996).

\bibitem{Melsen95} C. W. J. Beenakker, J. A. Melsen, and P. W. Brouwer,
Phys.\ Rev.\ B {\bf 51}, 13882 (1995).

\bibitem{Mel88a} P. A. Mello, E. Akkermans, and B. Shapiro, Phys.\
Rev.\ Lett.\ {\bf 61}, 459 (1988).

\bibitem{Bergmann} G. Bergmann, Phys.\ Rep.\ {\bf 107}, 1 (1984).

\bibitem{DeGennes} P. G. de Gennes, {\it Superconductivity of Metals and
Alloys} (Benjamin, New York, 1966).

\bibitem{ArgamanZee} N. Argaman and A. Zee, Phys.\ Rev.\ B
    {\bf 54}, 7406 (1996).

\bibitem{Samuel} S. Samuel, J.\ Math.\ Phys.\ {\bf 21}, 2695 (1980).

\bibitem{MelloSeligman} P. A. Mello and T. H. Seligman,
    Nucl.\ Phys.\ A {\bf 344}, 489 (1980).

\bibitem{MelloStone} P. A. Mello and A. D. Stone, Phys. Rev. B
    {\bf 44}, 3559 (1991).

\end{references}
\end{document}